\documentclass[11pt,a4paper]{article}

\pdfoutput=1 
\usepackage{jheppub}
\usepackage{amssymb}
\usepackage{dsfont}
\usepackage{color}
\usepackage{empheq}
\usepackage{slashed}

\def\bsub{\begin{subequations}}
\def\esub{\end{subequations}}

\newcommand*\widefbox[1]{\fbox{\hspace{2em}#1\hspace{2em}}}

\newcommand{\Gg}{\boldsymbol{\mathfrak{G}}}
\newcommand{\func}{\boldsymbol \Lambda}
\newcommand{\bphi}{\func}

\newcommand{\beps}{\boldsymbol{\epsilon}}

\newcommand{\vv}[1]{  \mbf v_{#1}}
\newcommand{\cO}{\mathcal O}
\newcommand{\df}{\Delta_\phi}
\newcommand{\reef}[1]{(\ref{#1})}
\newcommand{\Ds}{\Delta_*}

\newcommand{\dels}{\Delta_*}
\newcommand{\dgap}{\Delta_{\mbox{\tiny gap}}}

\newcommand{\be}{\begin{equation}}
\newcommand{\ee}{\end{equation}}
\newcommand{\bea}{\begin{eqnarray}}
\newcommand{\eea}{\end{eqnarray}}

\newcommand{\ud}{\mathrm d}
\newcommand{\mbf}{\mathbf}

\title{Extremal bootstrapping: go with the flow}

\author[a]{Sheer El-Showk}
\author[b]{Miguel F. Paulos}

\affiliation[a]{Laboratoire de Physique Th\'eorique et Hautes Energies, CNRS UMR 7589, Universit\'e Pierre et Marie
Curie, 4 place Jussieu, 75252 Paris Cedex 05, France}
\affiliation[b]{Theoretical Physics Department, CERN, Geneva, Switzerland}

\preprint{CERN-TH-2016-125}
\abstract{The extremal functional method determines approximate solutions to the constraints of crossing symmetry, which saturate bounds on the space of unitary CFTs. We show that such solutions are characterized by extremality conditions, which may be used to flow  continuously along the boundaries of parameter space. Along the flow there is generically no further need for optimization, which dramatically reduces computational requirements, bringing calculations from the realm of computing clusters to laptops. Conceptually, extremality sheds light on possible ways to bootstrap without positivity, extending the method to non-unitary theories, and implies that theories saturating bounds, and especially those sitting at kinks, have unusually sparse spectra. We discuss several applications, including the first high-precision bootstrap of a non-unitary CFT.}

\begin{document}
\maketitle

\section{Introduction}

The conformal bootstrap allows us to determine rigorous bounds on the parameter space of conformal field theories (CFTs) \cite{Polyakov:1974gs,Rattazzi:2008pe,Rychkov:2009ij,Caracciolo:2009bx,ElShowk:2012ht,El-Showk2014a,Kos:2014bka}. The bounds are possible thanks to the basic ingredients of unitarity and a convergent Operator Product Expansion (OPE) \cite{Wilson:1969zs}. When combined, these two properties allow for the decomposition of correlation functions into rapidly converging sums of positive terms \cite{Pappadopulo2012}, and from this positivity it is natural to expect bounds. For instance, one could imagine moving along a direction in parameter space along which some of the terms decrease. By positivity, they can become at most zero, and so the sums will contain fewer and fewer terms, until eventually we reach some minimum number. At this point we must stop: we have hit a boundary of parameter space, and to go any further would require us to relinquish positivity and with it unitarity.

The positive terms represent contributions to the correlation function from operators being exchanged. As we approach a boundary, the correlators receive contributions from fewer and fewer operators, which can be due to a sparser CFT spectrum, symmetries, or both. At the boundary itself, it is natural to expect that the correlators, and the CFTs themselves, should be extremely special. This logic goes some way in helping us understand the seemingly unreasonable effectiveness of the conformal bootstrap in accurately pinning down several theories of interest. 

So far most results have been numerical (though signficant analytic progress has been made by studying various limit, see e.g. \cite{Komargodski2013, Fitzpatrick2013}). One begins by reformulating crossing symmetry of conformal four point functions as linear or
semidefinite optimization problems which can be solved numerically. This allows us not only to rule out regions of CFT parameter space rigorously, but also to construct approximate solutions to crossing symmetry in certain cases.
In our previous work \cite{ElShowk:2012hu} we showed that in the {\em extremal} case, that is, on the boundary of the space of consistent solutions to crossing, these solutions are unique and can give excellent approximations to the low-lying spectrum of actual CFTs. In accordance with the expectations outlined in the previous paragraph, one does find that interesting CFTs tend to lie on the boundaries which allows us to extract their properties with great accuracy~\cite{El-Showk2014a}.

In this paper, we extend the philosophy and observations of
\cite{ElShowk:2012hu} and will examine in more detail what characterizes the CFTs that lie on the boundaries of parameter space. We call such CFTs extremal: they have sparse spectra, or more precisely, such theories contain correlation functions receiving contributions from as few operators as possible below any given cutoff in conformal dimension. We shall show that for these theories it is  possible to write down a set of extremality equations that fully characterize the solution to the crossing symmetry constraints. Perturbing these equations allows us to {\em flow} along the boundary of parameter space in a unique way. More specifically, given some ``seed'' extremal solution, we can then determine any other that is connected to it by a continuous variation by simply integrating a differential
equation. Furthermore, this method can be even used to derive the initial solution in the first place, as we shall see by explicit examples.  


To appreciate the power of this result it should be noted that
recent works on the bootstrap are often run on large computing clusters which can consume several years of CPU time. Each point along a boundary of a typical exclusion plot in parameter space must be independently computed via an expensive optimization step. Using our method a single point can be used to generate the entire plot within hours on a single laptop.


The outline of this paper is as follows. Below we begin with a condensed summary of our results to help orient the reader.  Then, in the next section, we provide a review of our previous work \cite{ElShowk:2012hu}, and present the extremal flow philosophy in a simple context. A more systematic development of the formalism is made in section \ref{sec:ext}, where we define extremal solutions to crossing as those satisfying certain Karush-Kuhn-Tucker optimality conditions in linear semi-infinite programming. In section \ref{sec:extflow} these conditions are perturbed to derive linearized flow equations, which lead to locally unique, extremal solutions to crossing in a neighbourhood of a given solution. Section \ref{sec:app1d} applies the flow equations in the simple context of conformal bootstrap in one-dimensional CFTs. 
The flow equations allow us to rederive the results of the usual bootstrap algorithms, at a small fraction of the computational cost. As examples we consider gap and OPE maximization, as well as flows that interpolate between these two cases. 

Section \ref{sec:disc} is reserved for an extended discussion of several issues. 
We connect our approach with the determinant method \cite{Gliozzi2013} and argue that the extremality equations make sense even without positivity.
We also consider singularities which may arise during flows. These singularities are a consequence of imposing positivity and usually signal interesting solutions to crossing. We briefly show how they can be resolved in general, and discuss a concrete example in one dimension.
%
Finally, we consider the question of convergence of the approximate crossing solutions that we construct. They are complemented by results on convergence of differentiated OPE expansions in appendix~\ref{sec:errder}. We finish this paper with some brief conclusions and an outlook on future work.

\subsection{Brief Summary of Results }

We provide here, for the reader's benefit, a condensed summary of the main results of this paper, highlighting our most important findings.
\begin{itemize}
\item We will formulate extremality conditions that characterize correlators at the boundary of parameter space and show how to solve linearized deformations of these conditions.  We focus our attention on deformations coming from varying the dimension of a scalar whose correlator we are considering but, in principle, this method can be applied to any differentiable parameter entering into the problem (e.g. the spacetime dimension, the rank of global symmetry groups, etc\ldots).

\item The same idea can be used to correct an approximately extremal solution to very high precision.  In practical applications this {\em error-correction} step is essential as solutions to the linearized deformation are always inexact. Perhaps most remarkably, we show that, in some cases, it may be possible to entirely do away with linear or semi-definite programming.  Using error-correction we can take a solution involving fewer operators and constraints and {\em upgrade} it to one involving more.  
We show that this works very well in $D=1$: starting from a single randomly guessed operator we have obtained extremal solutions containing up to 75 operators (which amounts to a 150 component truncation).

\item The methods above are vastly more computationally efficient than existing bootstrap techniques.  For instance, in $D=1$ we find that solving a standard bootstrap problem (such as maximizing an OPE coefficient) takes approximately 40 minutes (on a single CPU core).  Starting from this seed point, and using exactly the same parameters, we are able to {\em flow} to new points at a rate of approximately one point every 25 seconds -- a $\times100$ fold speedup!  This includes both the time to find a deformation of the original solution (e.g. a solution with some new value of the external scalar dimension) as well the time for several error correction steps to ensure that the new solution is extremal to very high accuracy.

\item While we have focused our first ``test-drive'' applications to $D=1$, the method itself is completely general and we have implemented and tested it in $D=2$ and $3$ as well.  What's more, it can be applied to the bootstrap of multiple correlators.

\item A final point is the observation that {\em extremality}, as defined above, is not strictly equivalent to positivity in the sense required by Linear and Semidefinite Programming. Indeed, the determinant method of \cite{Gliozzi2013} can be seen as a special case of our extremality approach. As a first application, we use our methods to bootstrap the generalized free fermion with negative conformal dimension to very high accuracy.
\end{itemize}

\section{Review: the extremal functional method} \label{sec:efm}

In this section we will briefly review the results obtained in reference \cite{ElShowk:2012hu}, as well as the basics of the numerical conformal bootstrap. We will show in a simple context the basics of extremality and flows, with the goal of developing an intuition for the more formal developments of the next section. We will be brief, so we direct the reader to the reference above as well as the reviews \cite{Qualls:2015qjb,Rychkov:2016iqz,Simmons-Duffin:2016gjk} for further information.

Consider the four-point correlator of a single scalar field $\phi$ with dimension $\Delta_\phi$ in a conformal field theory, 
	\bea
	\langle \phi (x_1)\phi (x_2)\phi (x_3)\phi (x_4)\rangle=\frac{g(u,v)}{x_{12}^{2\Delta_\phi}\,x_{34}^{2\Delta_\phi}},
	\eea
with $x_{ij}\equiv x_i-x_j$, and where $g(u,v)$ is a function of the conformally invariant cross-ratios
	\bea
	u=\frac{x_{12}^2\, x_{34}^2}{x_{13}^2\, x_{24}^2}, \quad v=\frac{x_{14}^2\,x_{23}^2}{x_{13}^2\,x_{24}^2}.
	\eea
The OPE  implies that the function $g(u,v)$ can be expanded in different channels. For instance, in the so-called direct channel we take $x_1\simeq x_2$  and get
	\bea
	g(u,v)=1+\sum_{\Delta,l} (\lambda_{\Delta,l})^2 \, G_{\Delta,l}(u,v). \label{crossing}
	\eea
The sum is over primaries of the conformal group, which are labeled here by two quantum numbers, their conformal dimension $\Delta$ and (traceless, symmetric) spin $l$.  Each primary together with its descendants contribute a conformal block $G_{\Delta,l}(u,v)$, \cite{DO1,DO2,DO3} to the correlator, weighed by the numbers $\lambda_{\Delta,l}$, which are the OPE coefficients appearing in the three-point function $\langle \phi \phi \mathcal O_{\Delta,l}\rangle$. The 1 in the sum above is the contribution of the conformal block of the identity which always appears in the four-point function of identical scalars. Equivalence of the expansion in the direct and crossed channels (where $x_1 \simeq x_4$) can be phrased as the non-trivial identity 
\bea	
\sum_{\Delta,l} (\lambda_{\Delta,l})^2\, F^{(\phi)}_{\Delta,l}(u,v)=-F^{(\phi)}_{0,0} \label{cross2}
\eea
with
\begin{eqnarray*}
	F^{(\phi)}_{\Delta,l}(u,v)\equiv v^{\Delta_\phi}\, G_{\Delta,l}(u,v)-u^{\Delta_\phi}\, G_{\Delta,l}(v,u).
\end{eqnarray*}
The crossing equations \reef{cross2} give an infinite, continuous set of constraints on hypothetical spectra and OPE coefficients of a CFT. We want to extract information from this equation, which can be done by truncating the constraints to a finite discrete subset. For instance we can Taylor expand each $F^{(\phi)}_{\Delta,l}$ to some finite order around a chosen $u,v$ point (usually $u=v=1/4$). We write the {\em truncated} crossing equations as
\begin{equation}\label{crossvec}
	\sum_i a_i \vv i = \mbf T, \qquad \mbf T=-\vv 0.
\end{equation} 
with $\vv i$ vectors of length $N$.
Here we have defined the {\em target} of the sum rule, $\mbf T$, which in this case is simply related to the contribution of the identity vector $\vv 0$. The vectors $\vv i$ are made up of derivatives of $F^{(\phi)}_{i}(u,v)$, the composite label $i=(\Delta, l)$ is
actually continuous (because $\Delta$ is), and the coefficients
$a_i=\lambda_i^2 \geq 0$ are positive for unitary CFTs.
 
The rapid convergence \cite{Pappadopulo2012} of the OPE ensures that solutions to (\ref{crossvec}) approximate quite well solutions to (\ref{cross2}) for even small values of $N$ (note that here $N$ is essentially the number of terms we keep from the Taylor expansion). To make rigorous statements, one generally attempts to prove that (\ref{crossvec}) has {\em no solutions}, since this implies that the same is true for \reef{cross2}. Of course we know that solutions to the full crossing constraints do exist so the non-trivial statement to be made will be that no solution can be found when one imposes restrictions on the spectrum. 

\begin{figure}
\begin{center}
\begin{tabular}{cc}
\includegraphics[width=7cm]{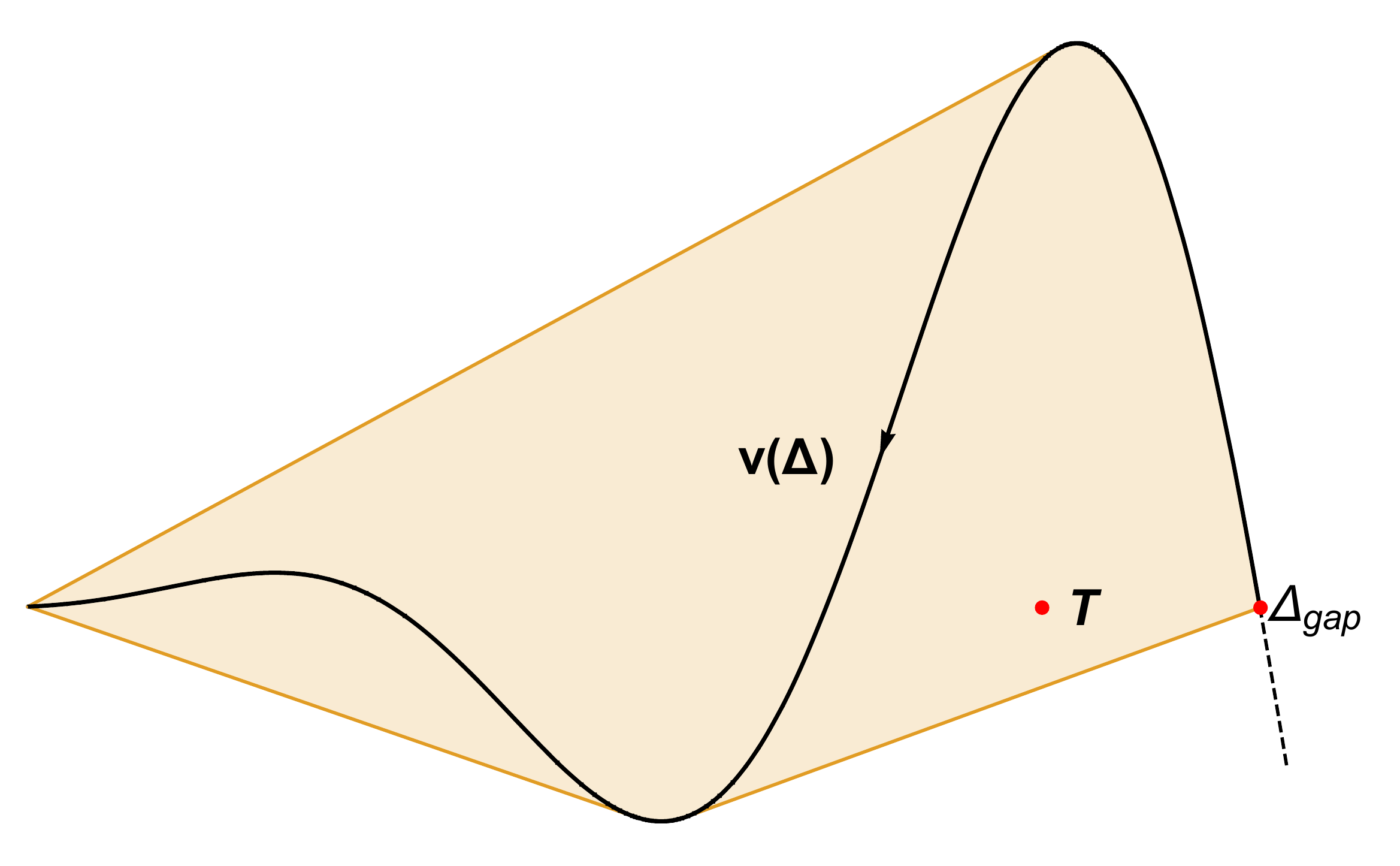}&
\includegraphics[width=7cm]{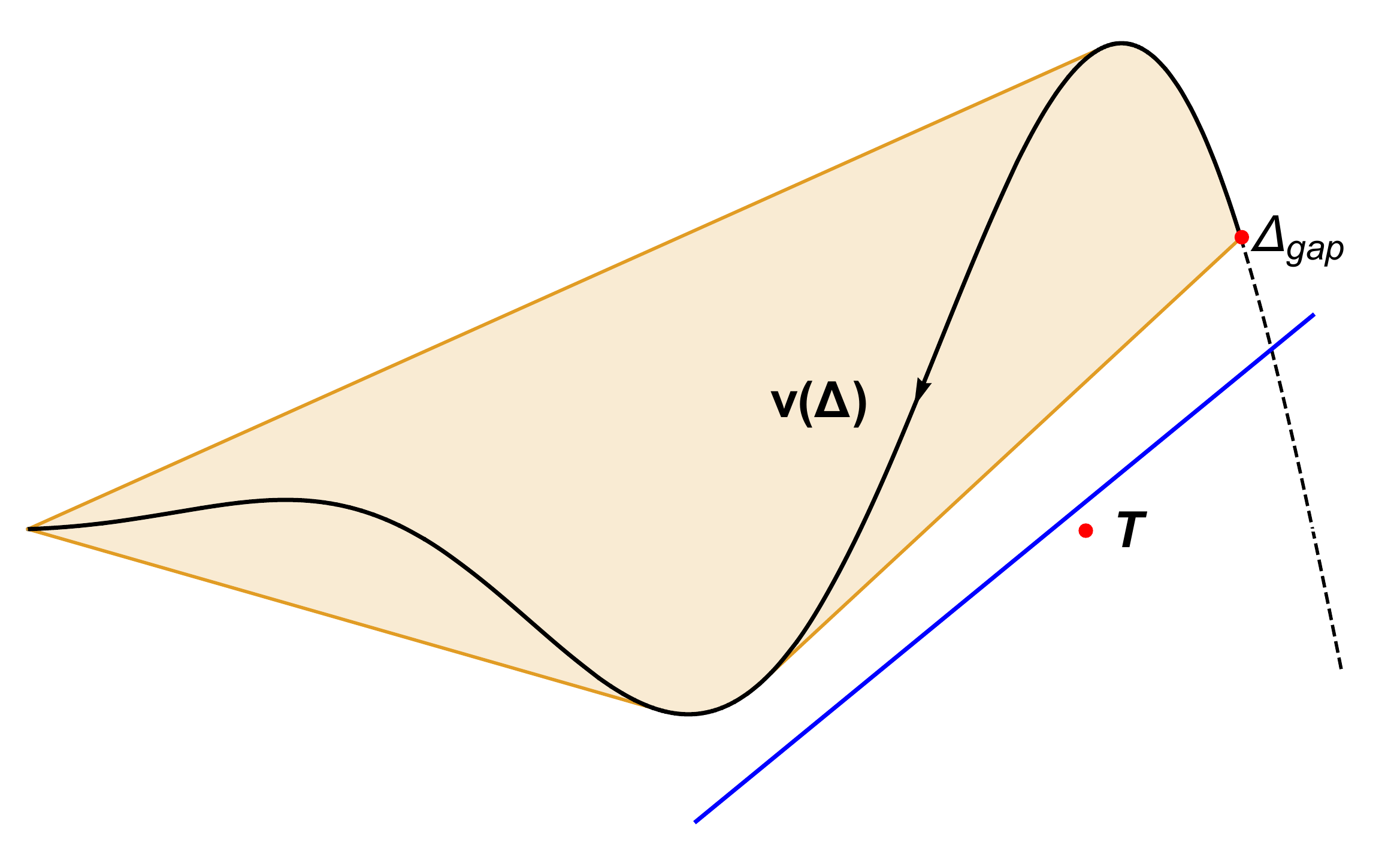}
\end{tabular}
\caption{Feasible vs unfeasible constraints. In black, the line of vectors labeled by the conformal dimension $\Delta$. The convex hull $H$ of these vectors includes the target $T$ on the left hand side. On the right it does not, and one can find a linear functional (the blue line) which separates the target from the remaining vectors.}
\label{fig:FeasVsUnfeas} 
\end{center}
\end{figure}

\begin{figure}
\begin{center}
\begin{tabular}{cc}
\includegraphics[width=7cm]{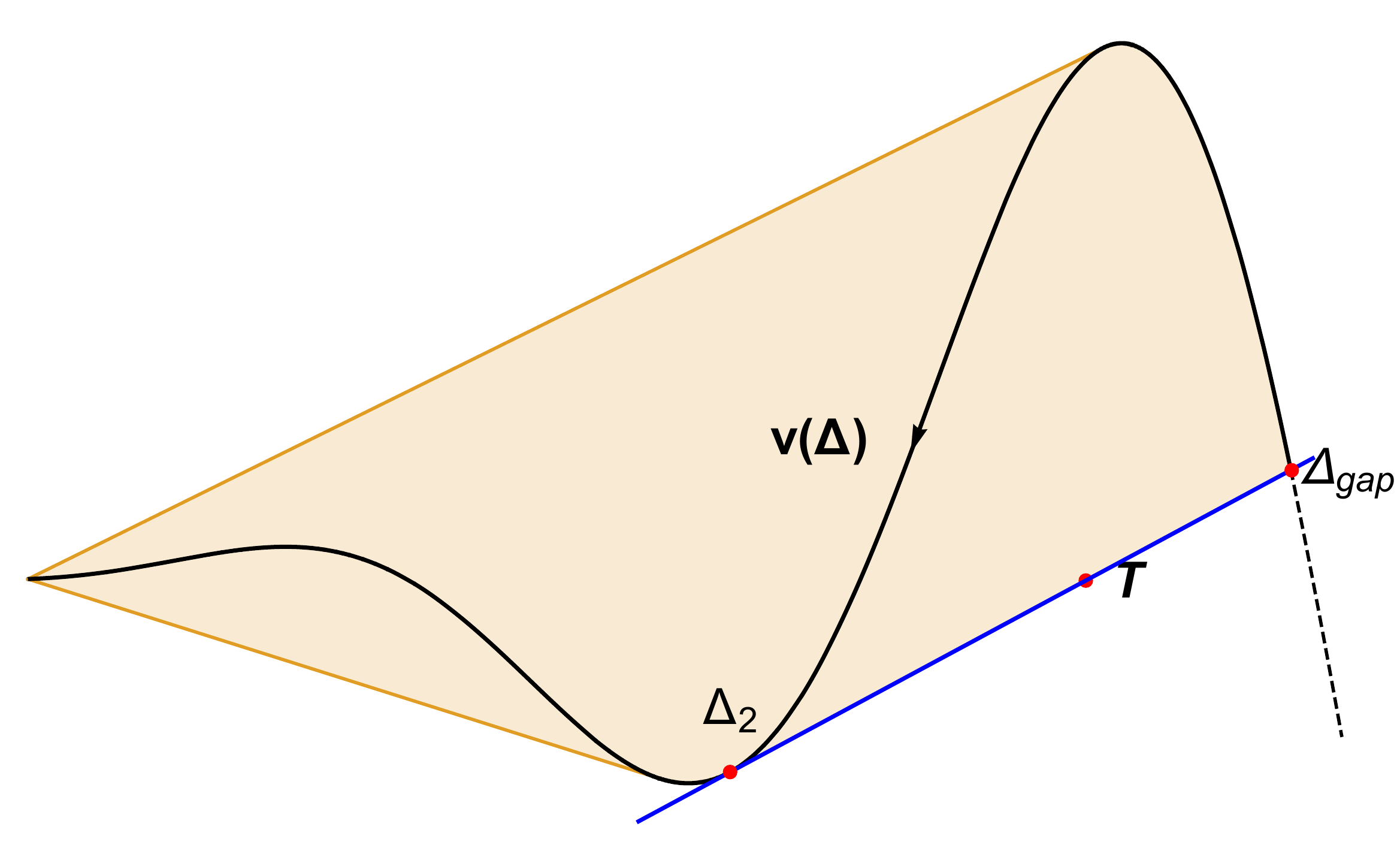}&
\includegraphics[width=7.5cm]{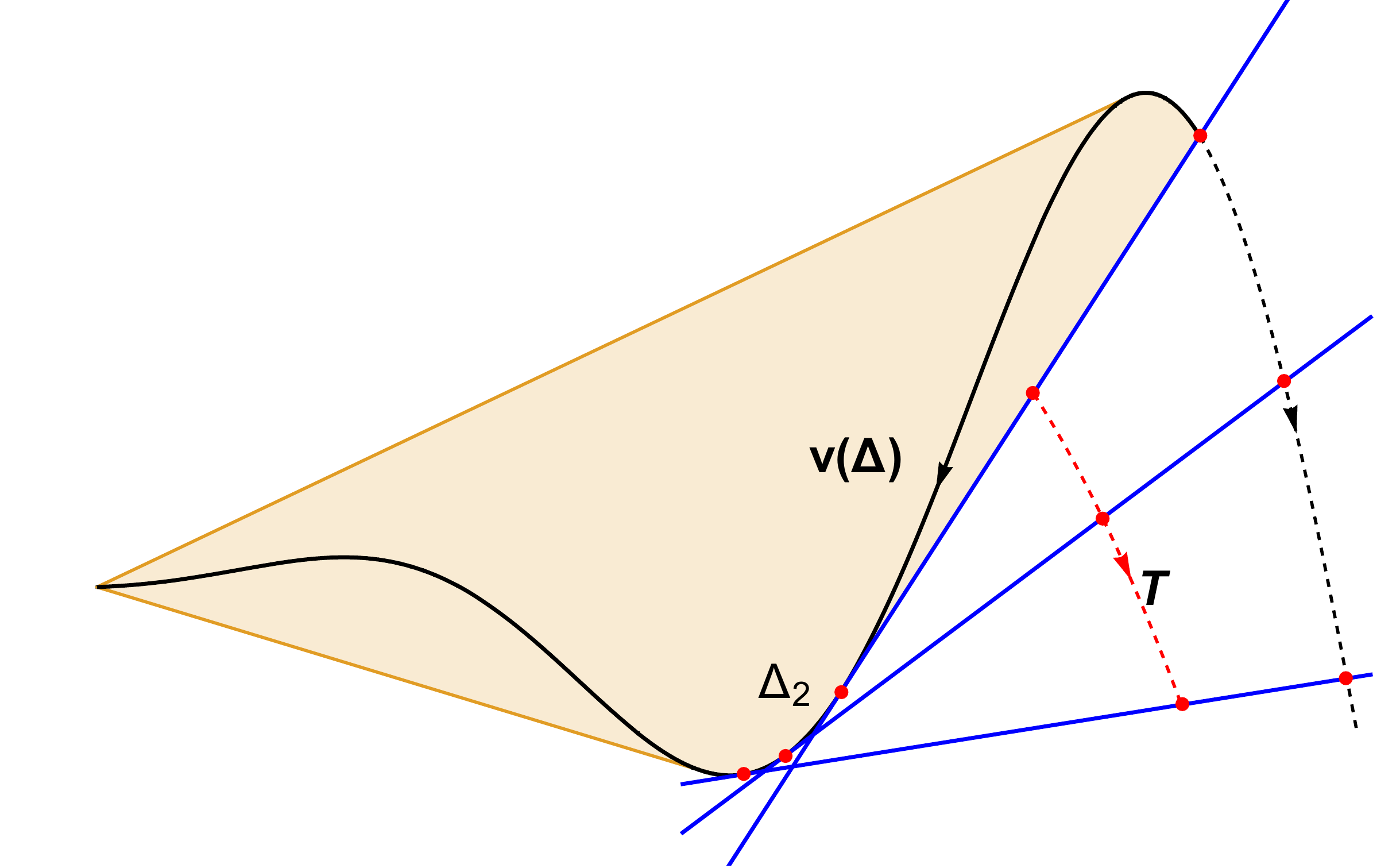}
\end{tabular}
\caption{On the left, the extremal case. The functional overlaps with a face of the convex hull of the vectors. On the right, we show how varying $\mbf T$ smoothly, the extremal functional varies continuously, keeping tangent to $H$.} 
\label{fig:extremalandflow}
\end{center}
\end{figure}

For instance, we may set a gap, $\Delta_{\mbox{\tiny gap}}$, to the first non-trivial scalar operator $\phi^2$ in the OPE $\phi \times \phi$, and increase it until a solution no longer exists. The allowed spectrum would take the form
\bea
S=\left\{ (\Delta,l): \left. \begin{array}{clc}
\Delta\geq &\Delta_{\mbox{\tiny gap}}, & l=0\\
\Delta\geq &d-2+l, & l>0
\end{array}\right.
\right\},
\eea
where the bounds for $l>0$ follow from unitarity. It is useful to consider the geometry associated with equation (\ref{crossvec}).  The vectors on the lefthand side come in continuous families, one for each $l$, parameterized by $\Delta$. To see whether a solution to the equation exists, we have to take the set all possible {\em positive} linear combinations of such vectors and check if $\mbf T=-\vv 0$ is among them. This set is a cone in $\mathds R^N$. In figure  \ref{fig:FeasVsUnfeas} we give a schematic view of the base of the cone when $N=3$. The base is the convex hull $H$ of the vectors $\vv i$, and we consider for simplicity a single family of vectors ({\em i.e.} one spin).
In the figure we show two distinct cases, corresponding to different gaps. It is clear that varying the gap can indeed lead to the absence of a solution to crossing. It also shows that when this is the case, one may prove it by finding a hyperplane (in this case a plane, which shows up as a line in the figure) which separates the target $\mbf T$ from all remaining vectors. Algebraically, we want $\func$ such that:
\begin{equation}\label{funceq}
\func \cdot \mbf T < 0, \qquad \func \cdot \vv i \geq 0 \quad \forall\, i \in S.
\end{equation}
Clearly if (\ref{funceq}) holds then
(\ref{crossvec}) cannot. If we continuously vary $S$ by increasing or decreasing $\Delta_{\mbox{\tiny gap}}$ we will find a transition point, corresponding to the maximal allowed gap. This is shown on the lefthand side of figure \ref{fig:extremalandflow}.
In this {\em extremal} case, $\func\cdot \mbf T\to 0$ while at the same time $\func \cdot \vv i=0$ for some number
of $\vv i$ (and $\Lambda\cdot \mbf v>0$ for all others). In the figure, this is true for the vector sitting at the gap, $\vv 1$, and some other vector $\vv 2$ with dimension $\Delta_2$. Since $\vv 1, \vv 2, \mbf T$ all lie on the same hyperplane, they must be linearly dependent, and hence we conclude that%
\bea
a_1 \vv 1+a_2 \vv 2=\mbf T \label{eq:cross2v}
\eea
for some positive\footnote{Positivity follows from $\mbf T$ being inside the convex hull of $\vv 1$, $\vv 2$.} $a_1, a_2$. 
Notice that we started off with three linear equations ($N=3$) but we ended up with a solution involving only two vectors. This is the first signature of extremality. This reduction signals the existence of the functional $\func$, which can be constructed explicitly as
\bea
\bphi_a\propto \epsilon_{abc} \vv 1^b \vv 2^c. \label{eq:lambda}
\eea
The sign of the proportionality constant in \reef{eq:lambda} can be fixed by demanding positivity of $\func$. Since $\func\cdot \vv 1=0$, a necessary condition for positivity is $\func \cdot \partial_\Delta \vv 1>0$. Having set up positivity of $\func$ in the neighbourhood of $\vv 1$, we can guarantee positivity everywhere by ensuring that all subsequent zeros of $\func$ are double zeros.
Geometrically, this means that the functional must be {\em tangent} to those vectors it touches, if they are in the interior of set $S$. In this case, there is only one such vector, $\vv 2$, and hence we must have
\bea
\bphi\cdot \partial_\Delta \vv 2=0  \label{eq:tangsimp}.
\eea
Now we notice the following: defining the functional via \reef{eq:lambda}, we see that equations (\ref{eq:cross2v}), (\ref{eq:tangsimp}) give $3+1=4$ conditions  for $2+2$ variables, namely the dimensions and OPE coefficients of the vectors $\vv 1,\vv 2$. Hence, we could in principle use these equations to directly find a solution to the crossing equations. Even in this simple example, this would involve finding solutions to complicated non-linear equations in four variables. However, once a given solution is found, it is trivial to linearize the equations to find how they change under smooth deformations. On the righthand side of figure \ref{fig:extremalandflow} we show one possibility. As we continuously vary the target $\mbf T$, the unique solution can be found by ``rolling'' the functional along the convex hull $H$. 

As a simple application, we can perturb equations \reef{eq:cross2v} and dot them with the unperturbed functional $\bphi$ to find the relation
\bea
\delta \Delta_1= \frac{\bphi \cdot \delta \mbf T}{\bphi\cdot \partial_\Delta \vv 1}.
\eea
Incidentally, this relation is correct even for more complicated setups as we will see in the following section. For now, we just note that the variations of other parameters may be determined analogously, essentially by solving a set of linear equations.

In the next section, we shall generalize considerably these observations. The main ideas however, are already here: extremal solutions to crossing constraints have functionals associated to them, which satisfy positivity and tangency conditions. These conditions determine in a unique way new solutions when we smoothly vary some parameter.

\section{The extremality equations}\label{sec:ext}
We consider a unitary CFT and examine a set of four point functions of an ensemble of operators labeled $\Psi_a$. Applying the OPE on some or all possible pairings of operators we are led to a set of constraints which are quadratic in the OPE coefficients. Quite generally these constraints take the schematic form:
\bea
\sum_\cO  \boldsymbol {\lambda}_\cO\cdot\Gg_{\cO,\Psi_a}(u,v) \cdot \boldsymbol{\lambda}_\cO=0.\label{eq:crossing}
\eea
Here $\boldsymbol{\lambda}_\cO$ stands for a vector of those OPE coefficients that are associated with the exchange of the conformal primary  operator $\cO$ in one or more correlation functions, and the conformal partial wave $\Gg_{\cO,\Psi_a}$ is a vector of matrices, whose elements are functions of the usual conformal cross-ratios $u,v$. This object is fully determined by conformal symmetry and group theory, and depends not only on the quantum numbers of operator $\cO$ but also on those of the operators $\Psi_a$ whose correlation functions we are considering. We should think of this equation as the generalization of \reef{cross2} in the previous section, with $\Gg_{\cO,\Psi_a}$ generalizing the functions $F_{\Delta,l}^{(\phi)}(u,v)$. We are interested in general solutions to these equations  and hence we should not really write a summation sign, since the set of quantum numbers of $\cO$ includes at least one continuous label, namely its conformal dimension $\Delta_\cO$.   

For our purposes it is convenient to move from the matrix form of these equations to something more reminiscent of (\ref{cross2}) by introducing auxiliary angular variables. We can do this by writing
\bea
\boldsymbol {\lambda}_\cO=\lambda_\cO\, \mbf n_\cO, \qquad \mbf n_\cO\cdot \mbf n_{\cO}=1.
\eea
The crossing equations now read
\bea
\sum_\cO  \lambda_\cO^2 \,\mbf F_{\cO,\Psi_a}(u,v) =0,\qquad   \mbf F_{\cO,\Psi_a}(u,v) \equiv \mbf{n_\cO}\cdot \Gg_{\cO,\Psi_a}(u,v)  \cdot \mbf{n_\cO} \label{eq:crossing2}.
\eea
Effectively, the operators $\cO$ now carry extra continuous ``quantum numbers'' given by $\mbf n_{\cO}$ which provide their orientation in OPE space. In the sum over operators we must now also sum over these new quantum numbers.   This effectively reduces the quadratic problem (\ref{eq:crossing}) to a ``linear'' one of the form (\ref{eq:crossing2}) (albeit with ``effective'' blocks depending on more continuous parameters).

At the cost of adding these parameters, we have managed to rewrite the full set of crossing equations in essentially the same form as one would the simple case \reef{cross2}, and their analysis will be similar. We first truncate the continuous set of constraints to a finite number $N$ of them, say by Taylor expansion,  and then make assumptions on which kinds of operators $\cO$ are allowed in the crossing relation to try to derive a contradiction. That is, we ask:
\begin{itemize}
\item[] {\bf Feasibility problem for region $\bar S$ (primal formulation):}
\bea
\exists\,\lambda^2_\cO:\qquad \sum_{\cO \in \bar S} \lambda^2_\cO\, \mbf F_{\cO,\Psi_a}=0, \qquad \lambda_{\mathds 1}^2=1 \label{eq:primalFeas}
\eea
\end{itemize}
where we have explicitly demanded that the identity operator contributes with unit coefficient. Typically the set $\bar S$ will be very large, involving several disconnected, continuous components. The crossing equations are now linear in the squares of the OPE coefficients, and since the set of vectors in $\bar S$ will generically contain a basis of $\mathds R^N$, this is only a non-trivial problem because we require $\lambda_\cO^2\geq 0$ by unitarity.

Rather than trying to solve the equations directly, we can attempt to rule out given sets $\bar S$ by constructing {\em positive linear functionals}. This leads us to the dual formulation of the feasibility problem:
\begin{itemize}
\item[] {\bf Feasibility problem for region $\bar S$ (dual formulation):}
\bea
\exists \func \in \mathds R^N: \qquad \func \cdot  \mbf F_{\cO,\Psi_a}\geq 0,\qquad  \forall \cO \in \bar S, \label{eq:dualFeas}
\eea
\end{itemize}
together with some normalization condition that prevents $\func$ from being identically zero, say $\func\cdot \mbf F_{\cO^*}=1$ for some particular $\cO^*$. If such a functional exists then we've ruled solutions to \reef{eq:crossing2} that contain only operators in subset $\bar S$. A typical application is to set a gap to the conformal dimension of the lowest lying operator with some set of quantum numbers, and increase it until a solution can no longer be found. If we cannot find solutions for a given set of constraints, we certainly won't able to find them by adding more, and therefore this procedure gives valid bounds for any truncation order.

At this point it is useful to rephrase the problem slightly. Firstly, given that we explicitly demanded the presence of the identity operator with unit coefficient, we can write
\bea
\sum_{\cO \in \bar S} \lambda^2_\cO\, \mbf F_{\cO,\Psi_a}=0\Leftrightarrow \sum_{\cO \in S} \lambda^2_\cO\, \mbf F_{\cO,\Psi_a}=-\mbf F_{\mathds 1,\Psi_a}
\equiv \mbf T
\eea
We have dropped the bar on $\bar S$ to indicate that the identity vector is no longer allowed in the sum. Also, we have denoted the contribution of the identity by $\mbf T$, the {\em target} of the sum rule. In general $\mbf T$ could be something else. For instance if we were interested in solutions where certain operators appeared with definite OPE coefficients we could move them onto the righthand side and incorporate them into $\mbf T$. Such operators wouldn't even need to come from set $S$ in principle. 

The second modification is to rewrite this feasibility test for region $S$ as a minimization problem. Let us introduce auxiliary $\mathds R^N$ vectors given by 
\bea
\mbf W^{(i)}=(\mbf W^{(i)}_{\ 1},\ldots, \mbf W^{(i)}_{N}), \qquad \mbf W^{(i)}_{\ j}=\delta_{ij} |\mbf T_i|
\eea
and write the sum rule as
\bea
\sum_{i=1}^N \mu_i \mbf W^{(i)}+\sum_{\cO \in S} \lambda^2_\cO\, \mbf F_{\cO,\Psi_a}=\mbf T.
\eea
with $\mu_i\geq 0$. Then the feasilibity problem for region $S$ is equivalent to 
\begin{itemize}
\item[] {\bf Feasibility problem for region $S$ (dual formulation):}
\bea
0\overset{?}{=}\underset{\func\neq 0}{\mbox{min}}\,  \func \cdot \mbf T:  \qquad \forall_{\cO \in S}\,\func \cdot  \mbf F_{\cO,\Psi_a}\geq 0, \qquad \forall_{i=1,\ldots,N}\, \func \cdot  \mbf W^{(i)}\geq -1. \label{eq:dualFeasOPE}
\eea
\end{itemize}
A further normalization condition should be added above to rule out $\bphi$ identically zero, for instance by strengthening one of the positive constraints to a strict inequality.
In this formulation, the minimum value will be reached when there is a solution containing a finite set of operators, for which $\bphi$ evaluates to zero, and a subset of the $\mbf W^{(i)}$, for which  $\bphi$ evaluates to minus one. Hence at the minimum we have $\bphi\cdot \mbf T=-\sum_i \mu_i\leq 0$, with zero being obtained if and only if $\mu_i=0$ for all $i$, giving us a solution without any auxiliary vectors.

This formulation is useful since it relates two problems that are usually presented as being disparate in the literature: feasibility (gap maximization) and OPE maximization (explained below). 
Suppose we already know that some region $S$ is feasible. Then a unique solution in $S$ may be singled out by asking:
\begin{itemize}
\item[] {\bf OPE maximization (primal formulation):}
\bea
\underset {\lambda^2_{\cO}}{\mbox{max}} \,\,C\equiv \sum_{\cO \in S} \lambda^2_\cO c_\cO: \qquad \sum_{\cO \in S} \lambda^2_\cO\, \mbf F_{\cO,\Psi_a}=\mbf T. \label{eq:primalOPE}
\eea
\end{itemize}
Once again, we can formulate the problem in the language of functionals:
\begin{itemize}
\item[] {\bf OPE maximization (dual formulation):}
\bea
\underset{\bphi\neq 0}{\mbox{min}}\,  \bphi \cdot \mbf T:  \qquad \forall_{\cO \in S}\,\bphi \cdot  \mbf F_{\cO,\Psi_a}\geq c_{\cO}  \label{eq:dualOPE}
\eea
\end{itemize}
In this dual formulation, the minimum value of this problem coincides with the maximum value of $C$. Notice that if $c_\cO\leq 0$ for all $\cO$ we need again an extra normalization condition to prevent $\bphi$ from being identically zero.
In principle the costs $c_\cO$ can be arbitrary function of the quantum numbers of $\cO$. In practice, we will simplify the discussion by restricting to costs which are zero almost everywhere except possibly for a few isolated vectors.

Since the feasibility test involves solving a problem of the type \reef{eq:dualOPE}, let us focus on the latter. In particular, we would like to obtain conditions characterizing the optimal solution. Luckily these are well known in the semi-infinite programming literature \cite{Hettich1993, Reemtsen1998} (\cite{Lopez2007} summarizes the main points). The problems written above are linear semi-infinite programs, and in this case second order optimality conditions reduce to the so-called Karush-Kuhn-Tucker conditions. The claim is that for an optimal solution there exists a set of $K\leq N$ vectors $\vv j$ (associated to $K$ operators as $\vv j\equiv  F_{\cO_j, \Psi_a}$, or possibly auxiliary vectors in the feasibility problem) and positive coefficients $a_j, (n_j^+)_a$ such that the following holds:
\begin{itemize}
\item[] {\bf Extremality conditions:}
\bsub
\begin{empheq}[box=\widefbox]{align}
Crossing &\qquad  &\sum_{j=1}^K a_j \vv j&=\mbf T,\label{eq:extreme1}\\
Saturation &\qquad & \func\cdot \vv j&=c_{\cO_j}, \label{eq:extreme2}\\
Tangency & \qquad & \func\cdot \nabla_a \vv j&=\left\{
\begin{tabular}{cc}
0 & if $\vv j \not\in \partial S$ \\
$(n_j^+)_a$ & if $\vv j \in \partial S$
\end{tabular}\right.\label{eq:extreme3}
\end{empheq}
\label{eq:extreme}
\esub

\end{itemize}
To make these conditions sufficient we must check that $\func$ has no negative regions, or more precisely, that $\func$ satisfies the constraints in \reef{eq:dualOPE}.

Let us analyse these conditions in more detail. The first are nothing but the existence of a feasible solution to the original, primal problem \reef{eq:primalOPE}, with the positive $a_j$ standing for the non-zero OPE coefficients. The second and third sets of conditions are more interesting. Firstly they tell us that all vectors in the solution to crossing must saturate the inequalities appearing in \reef{eq:dualOPE}. Secondly, they tell us that the functional must be {\em tangent} to those solution vectors which are on the inside of region $S$; or, if a vector lies on the boundary of the region, that the functional must be growing in the direction of the interior of $S$. That is the meaning of the quantities $n_j^+$ in those equations. For instance, if a neighbourhood $S_j$ of $\vv j\in \partial S$, looks locally like a patch of $\mathds R^N$ with coordinates $x_a$ satisfying  $x_1\geq 0,\ldots,x_k \geq 0 $ \ we would require $\bphi \cdot \frac{\partial \vv j}{\partial x_a}\geq 0$ for $a=1,\ldots, k$.

It is possible to give a geometric perspective on the extremality conditions, similar to the simple example in section \ref{sec:efm}. Let us forget about the overall scale set by $\mbf T$ and consider the problem projectively, as in that example.
The minimization problem \reef{eq:dualOPE} asks for a hyperplane ($\bphi)$ which should be as close as possible to the target $\mbf T$, while staying a finite (possibly zero) distance larger than $c_{\cO}$ away from the vectors $\mbf F_{\cO,\Psi_a}$.  Let us then imagine drawing some spheres with radius $c_{\cO}$ around these vectors.
The extremality conditions tell us that the hyperplane $\func$ closest to $\mbf T$ will necessarily touch some of these spheres. What's more, the tangency conditions imply that the hyperplane should be tangent to those spheres. This is easy to understand: if the tangency condition was not satisfied, than the hyperplane would cut through the interior of the sphere, leading to a violation of the inequalities in \reef{eq:dualOPE}. If some vectors lie on a boundary of the allowed space $S$, the tangency condition can be relaxed, since there are no other vectors along certain directions in the neighbourhood. In the region feasibility problem, where all costs $c_{\cO}$ are zero, the spheres simply shrink to zero size and the hyperplane now touches, and is tangent to, the convex hull of the vectors $\mbf F_{\cO, \Psi_a}$ (as near $\Delta_2$ in figure \ref{fig:extremalandflow}).

\section{Extremal flows} \label{sec:extflow}

The extremality equations generically define a locally unique solution to crossing. This uniqueness means that if we make some smooth deformation of the equations, we should be able to {\em flow} to a nearby extremal solution in a unique way. In this section we examine these flow equations in detail. For simplicity, we shall focus on the case where vectors are labeled by a single continuous parameter plus an undetermined set of other discrete labels. More parameters (such as the angles in eqn. (\ref{eq:crossing2})) can be handled straightforwardly, and we postpone their discussion to later work. We first discuss the OPE maximization problem \reef{eq:primalOPE}, \reef{eq:dualOPE} and only afterwards the region feasibility problem \reef{eq:primalFeas},\reef{eq:dualFeas}, since the latter is very simply related to the former.

\subsection{OPE maximization}
Consider the maximization of the OPE coefficient of a single operator, given by a vector $\vv 1$ with associated cost parameter $c_{\cO_1}=1$, and suppose we have found an initial solution, which must satisfy the extremality conditions \reef{eq:extreme}. It is convenient to split the $K\leq N$ vectors in the solution to crossing into several groups. Firstly, there is the vector $\vv 1$ whose OPE coefficient we are maximizing. We will assume that this vector doesn't move under deformations, {\em i.e.} its conformal dimension $\Delta_1$ is fixed in the formulation of the problem (or it changes in some prescribed way as we deform the problem). Next, we take the set of $n_f$ boundary vectors ($\vv j \in \partial S$), and label them by $\mbf f_1, \mbf f_2,\ldots,\mbf f_{n_f}$. The letter {\em f} stands for {\em fixed}, since as we shall see these vectors cannot move away from the boundary of $S$ under perturbations. The bulk vectors ($\vv j \not \in \partial S$) are further split into $n_d$ {\em doubles} $\mbf d_1,\ldots \mbf d_{n_d}$ and $n_s$ {\em singles} $\mbf s_,\ldots, \mbf s_{n_s}$. The precise way in which this split is made is immaterial. The only constraint is that we should take $n_d$  as large as possible, while satisfying
\bea
K=1+n_f+n_s+n_d,\qquad N=1+n_f+n_s+2\,n_d.
\eea
The reason for the second constraint is as follows. Suppose we had a solution with $K=N$. Then we must have $n_d=0$, and we could simply define the functional as
\bea
\func(\bullet)=\frac{\langle \mbf f_1\,\ldots \mbf f_{n_f}\,\mbf s_1\,\ldots \mbf s_{n_s} \bullet\rangle}{\langle \mbf f_1\,\ldots \mbf f_{n_f}\,\mbf s_1\,\ldots \mbf s_{n_s}\,\vv 1\rangle} 
\eea
where the notation $\langle \mbf v_1 \ldots \mbf v_{N}\rangle$ means a contraction of the totally antisymmetric Levi-Civita symbol with the corresponding vectors, and in particular it is the determinant of the matrix whose columns are $\mbf v_1, \ldots \mbf v_n$. This definition automatically solves the saturation conditions \reef{eq:extreme2}, {\em i.e.} $\func(\vv j)=0$ for all vectors in the solution. However, the remaining conditions in \reef{eq:extreme3} are non-trivial. Now suppose $K<N$. Then in our definition of the functional there will be some room to be filled in the determinants. We can use this extra room to also solve some of the tangency conditions automatically:
\bea
\func(\bullet)=\frac{\langle \mbf f_1\,\ldots \mbf f_{n_f}\,\mbf s_1\,\ldots \mbf s_{n_s}\, \mbf d_1\,\partial_\Delta \mbf d_1\ldots \mbf d_{n_d}\,\partial_\Delta \mbf d_{n_d}\, \bullet\rangle}
{\langle \mbf f_1\,\ldots \mbf f_{n_f}\,\mbf s_1\,\ldots \mbf s_{n_s}\, \mbf d_1\,\partial_\Delta \mbf d_1\ldots \mbf d_{n_d}\,\partial_\Delta \mbf d_{n_d}\, \vv 1\rangle} \label{eq:func}
\eea
As $K$ decreases for fixed $N$, so does the number of singles, and the smaller the number of tangency conditions we will have to satisfy. In particular,  if $n_s=0$ all the tangency conditions in \reef{eq:extreme3} for bulk vectors are automatically satisfied.

With these preliminaries sorted out, let us now see what we can say about perturbations to the solution. For instance, one possibility is to deform the target $\mbf T$. In fact, this is the most general case, since any perturbation at all can always be moved to the righthand side of the crossing equations by redefining $\mbf T$. For a small deformation, we can assume that the discrete labels of the vectors do not change\footnote{There could of course be discontinuities in these discrete labels at special points in parameter space. We shall discuss these singularities in more detail in section \ref{sec:sings}.}, but the continuous ones can and do. Here the only such parameters are the conformal dimension $\Delta_i$ and the OPE coefficients $a_i$. Recalling that the dimension of $\vv 1$ stays fixed, the counting of degrees of freedom gives %
\bea
\#\mbox{d.o.f.}=2K-1=N+n_s+n_f.
\eea
At the same time, the extremality conditions give $N$ crossing equations and $n_s$ extra tangency constraints. This is because the saturation conditions are automatically satisfied by our definition of the functional (cf. eqn.~\reef{eq:func}), and the tangency conditions for boundary vectors should automatically remain true under small perturbations (since the $n_a^+$ are finite positive numbers to begin with it). So, overall it seems we have a mismatch by $n_f$ in the number of constraints vs degrees of freedom. The solution, as we mentioned above, is that the fixed vectors must remain, well, fixed, which means they actually only contribute one degree of freedom each, namely their respective OPE coefficient. 

Let us see why this should be the case explicitly from the extremality equations, together with overall positivity of $\func$. By varying the crossing sum rule one finds
\bea
\sum_{i=1}^K\left( \delta a_i  \vv i+ \delta \Delta_i a_i \partial_\Delta  \vv i\right)= \delta \mbf T. \label{eq:varcross}
\eea
Acting with the unperturbed functional defined in \reef{eq:func} this becomes
\bea
\delta a_1+\sum_{k=1}^{n_f} a_{f_k}\delta \Delta_{f_k} \bphi\cdot \partial_\Delta \mbf f_k&=&\bphi \cdot \delta \mbf T \nonumber \\
\Leftrightarrow \bphi \cdot \delta \mbf T - \sum_{k=1}^{n_f}\,a_{f_k}\, \delta \Delta_{f_k} n^+_k&=&\delta a_1 \label{eq:prea1var}
\eea
where $n_k^+$ are positive/negative if $\mbf f_k$ sits on a left/right boundary. Now, if after variation the new functional is to be positive everywhere in $S$, we must have $\delta \Delta_{f_k} n^+_k \geq 0$. But since we are maximizing $a_1$ we see that the maximum value is then obtained when all $\delta \Delta_{f_k}=0$, which was what we wanted to show. %

Going back to our counting of degrees of freedom, we have
\bea
\#\mbox{d.o.f}=N+n_s
\eea
The case $n_s=0$ is especially simple. Since then one needs only use the crossing equations \reef{eq:extreme1}, the set of linear equations \reef{eq:varcross} are sufficient to determine the perturbed solution. In the case where $n_s>0$ this is not enough, and we must also use the tangency conditions to obtain a unique solution. These conditions tell us that the gradients of the singles should be killed by the functional, 
\bea
\bphi\cdot \partial_\Delta \mbf s_k=0 \label{eq:sk}
\eea
In other words, $\partial_\Delta \mbf s_k$ must be linearly dependent on the vectors that span the hyperplane defined by $\bphi$. 
Defining the matrix $\boldsymbol{\mathcal A}$ as
\bea
\boldsymbol{\mathcal A}\equiv \Bigg(\vv 1\, \mbf f_1 \ldots \mbf f_{n_f}\, \mbf s_1\ldots \mbf s_{n_s}\, \mbf d_1\ldots \mbf d_{n_d} \,\partial_\Delta \mbf d_1\,\ldots \, \partial_\Delta \mbf d_{n_d} \Bigg),
\eea
and using the definition of the functional \reef{eq:func}, we can write \reef{eq:sk} as
\bea
(1\, 0 \ldots 0)\cdot \boldsymbol{\mathcal A^{-1}}\cdot \partial_\Delta \mbf s_k=0.
\eea
We now perturb these conditions to linear order, much as for \reef{eq:varcross}. Actually we will do the following. Firstly, we shall not assume that these conditions are exactly satisfied, so we will allow some quantity $S_k$ on the righthand side. Secondly, we simply say that this quantity gets perturbed in some way. For instance, varying some intrinsic parameter of the problem, such as the dimensions of the operators $\Psi_a$, will generate such a perturbation. We get
\bea
\bphi\cdot  \partial_\Delta \delta \mbf s_k+(1\, 0 \ldots 0)\cdot (\delta \boldsymbol{\mathcal A^{-1}})\cdot \partial_\Delta \mbf s_k=\delta S_k.
\eea
In applications it is cumbersome to work with $\delta \boldsymbol{\mathcal A^{-1}}$. However, we can use a trick:
\bea
(1\, 0 \ldots 0)\cdot (\delta \boldsymbol{\mathcal A^{-1}})=\bphi \cdot \boldsymbol{\mathcal A}\cdot (\delta \boldsymbol{\mathcal A^{-1}})=-\bphi \cdot \delta\boldsymbol{\mathcal A}\cdot \boldsymbol{\mathcal A^{-1}}.
\eea
The variation equations can now be written as:
\bea
(\bphi\cdot \partial_\Delta^2 \mbf s_k) \delta \Delta_{s_k}-
\sum_{j=1}^{n_s} \beta_{kj} 
\, S_j\, 
\delta \Delta_{s_j}
-
\sum_{p=1}^{n_d} \eta_{kp} (\bphi\cdot \partial_\Delta^2 \mbf d_p)\, \delta \Delta_{d_p}=\delta S_k, \label{eq:sheer}
\eea
where the various coefficients are given by
\bea
\boldsymbol{\mathcal A}^{-1}\cdot \partial_\Delta s_k=
\left(
\begin{tabular}{c}
$S_k$\\
$\alpha_{k}$\\
$\beta_{k}$\\
$\gamma_{k}$\\
$\eta_{k}$
\end{tabular}
\right).
\eea

Overall, the {\em extremal flow} equations \reef{eq:varcross} together with \reef{eq:sheer} provide us with $N+n_s$ equations which can be readily inverted for the $N+n_s$ parameters of the new solution. For instance, when $n_s=0$ we get
\bea
\left(
\begin{tabular}{c} 
$\delta a_i$\\\hline
$\delta \Delta_i$
\end{tabular}
\right)=\boldsymbol{\mathcal M}^{-1}\cdot\delta \mbf T, \label{eq:flowsimple}
\eea
with the matrix
\bea
\boldsymbol{\mathcal M}\equiv \Bigg(\vv 1\, \mbf f_1 \ldots \mbf f_{n_f}\,\,\mbf d_1\ldots \mbf d_{n_d} \,\Bigg|(a_{d_1} \partial_\Delta \mbf d_1)\,\ldots \, (a_{d_{n_d}} \partial_\Delta \mbf d_{n_d}) \Bigg).
\eea
As a particularly simple and nice application of these equations we can determine the variation of the OPE coefficient that is being maximized: 
\bea
\delta a_1=\bphi \cdot \delta \mbf T= (\boldsymbol{\mathcal M}^{-1}\cdot \delta \mbf T)_1.  \label{eq:opemax1}
\eea
This result holds even in the presence of singles, since it also follows from \reef{eq:prea1var}.

We will solve these flow equations numerically in section \ref{sec:app1d}. But before we do so, let us discuss flows for the feasibility problem.

\subsection{Gap maximization}
We have already shown that the feasibility problem for a certain region can be thought of as a special case of OPE maximization, where there are some extra auxiliary vectors in the sum rule. A region is feasible when the OPE coefficients of such vectors vanish altogether. Here we are interested in the limiting case where a given region is barely unfeasible. We expect that the solution should contain a single auxiliary vector with a very small OPE coefficient, which will tend to zero as the region becomes feasible. For definiteness, here we shall consider the {\em maximal gap} problem: given some assumptions on the spectrum, we attempt to maximize the conformal dimension of the first operator with some discrete quantum numbers. 

This problem is very similar to OPE maximization, with two differences. Firstly, the role of $\mbf v_1$ in the previous section is now played by an auxiliary vector. Secondly, its OPE coefficient should be zero along the flow. Now, this may seem odd, since in the previous section the extremality equations were sufficient to uniquely fix a solution under perturbations. Adding an extra constraint would seem to make our problem overdetermined. The solution of course is that the maximum allowed value for the gap must vary along the flow. In practice, there will be a hypothetically fixed vector which will not be able to stay fixed. This will be exactly the vector whose gap we are maximizing. 

This can be understood more clearly from the perturbed crossing equations 
\bea
\sum_{i=1}^K\left( \delta a_i  \vv i+ \delta \Delta_i a_i \partial_\Delta  \vv i\right)= \delta \mbf T \label{eq:varcross2},
\eea
where now $a_1=0$ and $\vv 1$ is some fixed auxiliary vector (i.e. for which $\partial_\Delta \vv 1=0$). When we act with the functional \reef{eq:func} as before  this becomes
\bea
\bphi \cdot \delta \mbf T - \sum_{k=1}^{n_f}\,a_{f_k}\, \delta \Delta_{f_k} n^+_k&=&\delta a_1=0
\eea
where we are demanding that the OPE coefficient of the auxiliary vector remains zero under perturbations. The only way that this equation can be satisfied is if one (or some linear combination) of the fixed vector dimensions is allowed to change. That is, in order to flow we must deform the region $S$, or more prosaically, vary the gap. In the simplest case, we vary a single vector and we obtain
\bea
\delta \Delta_{f_1}=\frac{\bphi \cdot \delta \mbf T}{\bphi \cdot \partial_{\Delta} \vv {f_1}} \label{eq:gapmax1}
\eea
To summarize, the loss of one degree of freedom (the OPE coefficient $a_1$) must be compensated by allowing one of the ``fixed'' vectors to move along the flow. The rest of the analysis now proceeds exactly as before.

\section{Applications in $D=1$} \label{sec:app1d}
In this section we will show how to apply the flow equations derived in the previous section to obtain fast, precise numerical bounds. Our setup will be the simplest possible: bootstrapping a single correlation function of identical scalar operators $\phi$ in a one-dimensional CFT. In one dimension there is no spin, which means that in the OPE $\phi\times \phi$ operators are classified solely by their conformal dimension. This means we do not have to worry about discrete labels, which can in principle jump discontinuously along the flow, as we'll discuss in section \ref{sec:sings}. Our goal here is mainly to show the potential of extremal flows to dramatically increase the computational efficiency of bootstrap methods, leaving more realistic applications for future work. 

Consider then bootstrapping the four point function
\bea
\langle \phi(x_1) \phi(x_2)\phi(x_3) \phi(x_4)\rangle=\frac{g(x)}{|x_{12}|^{2\Delta_\phi}|x_{34}|^{2\Delta_\phi}}
\eea
with $x$ the conformal cross-ratio $x=\frac{(x_1-x_2)(x_3-x_4)}{(x_1-x_3)(x_2-x_4)}$.
In the following we will consider two different bootstrap applications. The first is the problem of maximizing the gap to the first non-trivial scalar $\equiv \phi^2$ in the OPE $\phi\times \phi$. That is, we want:%
\begin{itemize}
\item {\bf Application 1}
\bea
\underset {\lambda^2_{\Delta}\geq 0}{\mbox{max}}\, \dgap:  \qquad \sum_{\Delta\geq \dgap} \lambda_\Delta^2 \mbf F_{\Delta,\Delta_\phi}=-\mbf F_{0,\Delta_\phi}, \label{eq:prob1}
\eea
\end{itemize}
The second will be to maximize a specific OPE coefficient, given a gap:
\begin{itemize}
\item {\bf Application 2}
\bea
\underset {\lambda^2_{\Delta}\geq 0}{\mbox{max}}\, \lambda^2_{\dels} : \qquad \lambda_{\dels}^2 F_{\dels,\Delta_\phi}+\sum_{\Delta\geq \dgap} \lambda^2_\Delta\, \mbf F_{\Delta,\Delta_\phi}=-\mbf F_{0,\Delta_\phi}. \label{eq:prob2}
\eea
\end{itemize}
For these problems we take
\bea
\mbf F_{\Delta,\Delta_\phi}=\left(\partial_x F^{(\phi)}_\Delta,\partial_x^3 F^{(\phi)}_\Delta,\ldots,\partial_x^{2N-1} F^{(\phi)}_\Delta\right )\bigg|_{x=1/2} \label{eq:Fders}
\eea
with
\bea
F_{\Delta}^{(\phi)}=(1-x)^{2\Delta_\phi} G_{\Delta}(x)-x^{2\Delta_\phi} G_{\Delta}(1-x)
\eea
and the $d=1$ conformal block
\bea
G_\Delta(x)\equiv x^\Delta \, _2F_1(\Delta,\Delta,2\Delta,x).
\eea
In both cases, an extremal solution is a collection of $K$ vectors $\vv i\equiv \mbf F_{\Delta_i,\Delta_\phi}$ and associated functional (of course the latter is completely fixed, up to a scale, by the vectors $\vv i$) satisfying the conditions \reef{eq:extreme}, with the target $\mbf T\equiv -\mbf F_{0,\Delta_\phi}$.

It will be useful below to point out that the first problem has been considered in the past \cite{Gaiotto2014}, and the bound was found to be nearly saturated by the so-called generalized free fermion CFT\footnote{While it may seem odd that a fermionic four-point function can be recovered from a supposedly scalar field bootstrap, one must remember that we are in one-dimension where there is no spin. We will not dwell on this further, pointing out only that if one considers only the OPE channels (12)(34) and (14)(23), there is no way to detect the fermionic nature of the operators.}. This is simply the theory of a fermion with non-canonical dimension $\Delta_\phi\neq 0$, and two-point function
\bea
\langle \phi(x_1)\phi(x_2)\rangle=\frac{\text{sign}(x_1-x_2)}{|x_1-x_2|^{2\Delta_\phi}},
\eea
with all other correlation functions of $\phi$ factorizing into products of two-point functions. In particular,%
\bea
\langle \phi(x_1) \phi(x_2)\phi(x_3) \phi(x_4)\rangle=\frac{\text{sign}(x_1-x_2)\text{sign}(x_3-x_4)}{|x_1-x_2|^{2\Delta_\phi}|x_3-x_4|^{2\Delta_\phi}}\, F(x)
\eea
with
\bea
F(x)&=&1+\left(\frac{x}{1-x}\right)^{2\Delta_\phi}-x^{2\Delta_\phi}\nonumber \\
&=&1+\sum_{j=0}^{+\infty}\, \lambda_j^2 G_{\Delta_j}(x) \nonumber \\
&=&1+\sum_{j=0}^{+\infty}\, \frac{2 \,(2 \Delta_\phi)^2_{2j+1} }{(2j+1)!\,(4\Delta_\phi+2j)_{2j+1}} G_{2\Delta_\phi+2j+1}(x),\label{eq:gff}
\eea
and $(a)_n$ is the Pochhammer symbol. This decomposition satisfies the crossing equations~\reef{cross2}.

\subsection{Error correction}
In the following we would like to start from a fixed solution to problems \reef{eq:prob1}, \reef{eq:prob2} and perturb them to flow to new ones. Before we flow, we must have an extremal solution to begin with, but we are immediately faced with the problem that in practice, the numerical linear or semidefinite programming algorithms that are usually used to solve these problems can only obtain approximately extremal solutions. In the case of OPE maximization this is not such a big problem. The algorithms proceed via a series of iterations which must be cut-off at some point. Usually this is done by demanding that the OPE coefficient that is being maximized has converged to within some amount which can be chosen very small in practice (say machine precision, $10^{-15}$). However, \reef{eq:prob1} is significantly more problematic. This is because in practice one must check, for several choices of $\dgap$, whether the crossing constraints have a solution, and for each such choice one must solve a separate OPE maximization problem. One typically resorts to a bisection scheme in $\dgap$ to find the maximal value, and hence an accuracy of $\epsilon$ in $\dgap$ requires order $-\log_2(\epsilon)$ separate OPE maximizations. A further hindrance is the fact that as one approaches the maximal value, the OPE maximization problem takes longer and longer to converge.

A final complication is that the flow equations are simply linearized approximations, which will inevitably introduce errors when we integrate them a finite distance. So even in the best case scenario where we would apply flows to solutions of the OPE maximization problem \reef{eq:prob2}, the error would either rapidly grow, or we would have to resort to using very small perturbations. Altogether these difficulties would seem to reduce extremal flows to very limited applications.

Fortunately for us, there is a simple solution to these problems. The point is that flows can be used not only to perturb away from a given solution, but also to improve an approximate solution. That is, extremal flows are in a sense {\em self-correcting}. The reason for this is simply that the extremality equations \reef{eq:extreme} are of the schematic form $f(x)=0$. Hence, given an approximate solution $x^{(p)}$, we can use Newton's method,
\bea
x^{(p)} \to x^{(p+1)}=x^{(p)} -\frac{f(x^{(p)} )}{f'(x^{(p)})}
\eea
 to rapidly find an improved one.
In practice this is extremely simple to implement: we merely consider a ``flow'' where in the linearized equations one sets as a source the error in the extremality equations, i.e. the failure to satisfy the crossing and tangency conditions. For instance, in the case where there are no singles, we can write (cf. \reef{eq:flowsimple})
\bea
\left(
\begin{tabular}{c} 
$\delta a_i^{(p+1)}$\\\hline
$\delta \Delta_i^{(p+1)}$
\end{tabular}
\right)
=\boldsymbol{\mathcal M}^{-1}(a_i^{(p)}, \Delta_i^{(p)})\cdot \delta \mbf T^{(p)} \label{eq:errcorr}
\eea
where $\delta \mbf T^{(p)}$ here stand for the failure to satisfy crossing, i.e.
\bea
\delta \mbf T^{(p)}=\sum_{i=1}^K a_i^{(p)} \mbf v_i^{(p)}-\mbf T, \qquad \mbf v_i^{(p)}\equiv \mbf v(\Delta_i^{(p)}).
\eea
This is not really a flow since there is no parameter varying continuously, but the equations used are exactly the same (we could introduce a fictitious coefficient, $\alpha$, in front of $\mbf T^{(p)}$ and then think of this as a flow from $\alpha=1$ to $\alpha=0$). We call this {\em error-correction}, and it is the key feature of the extremality equations which makes the extremal flow method both feasible and incredibly powerful.

We can use error-correction to improve any given approximate solution to crossing, obtained say by linear programming methods. This holds both for the OPE  and gap maximization problems. As explained above, for the latter this is extremely useful, since it allows us to quite easily obtain very accurate values for $\dgap$ without need for bisection, as we shall see in the applications below. We also use error correction to improve a solution after a flow. That is, after varying some continuous parameter such as $\Delta_\phi$ we can flow from one extremal solution to another, approximate one, which can then be systematically improved by using error correction. In any case, we have found that in practice, iterating \reef{eq:errcorr} a handful of times suffices to give us solutions which are extremal to within precision -- typically over a hundred digits.

One should note that error correction does not always work (just as Newton's method). If the initial solution is not sufficiently close to the exact one, then one may find that error correction diverges, leading to systematically worse solutions. In this case one may either attempt to get a better initial guess, or use a modified Newton's method with line search. In practice we have found that for higher values of $N$ (the truncation parameter) a better initial guess is required in order for error correction to work. A simple strategy is to initially flow in small steps, so that the linearized approximation is sufficient to get a good initial guess. Once some initial data is available, we can use it to extrapolate better guesses for other points (though this is not necessary and the method works quite well even without this).

\subsection{Upgrading}
\label{sec:upgrade}
Our first application is not a really a flow, but rather demonstrates the power of error correction on its own. We call it ``upgrading'', and it means generating an extremal solution with a higher value of the truncation parameter $N$ from a lower one. Recall that $N$ controls the accuracy of the bootstrap bounds, since higher values of $N$ include more crossing symmetry constraints. Of course, higher values are also more difficult to obtain numerically.

We will consider the gap maximization problem \reef{eq:prob1}, and set $N=2$ to begin with. For concreteness, we also choose $\Delta_\phi=0.3$. In this case, we expect the extremal solution to contain a single vector which will sit at the maximum allowed value of $\dgap$. To obtain this extremal solution we will not use linear programming at all: we simply make an initial guess and then do error correction. For instance, starting with an initial guess $\dgap=1$, $\lambda_{\dgap}^2=1$, we find after ten iterations a new solution with $\dgap\simeq 1.874, \lambda_{\dgap}^2\simeq 0.414$. This solution satisfies crossing to within one part in $10^{300}$. To check that it is the extremal solution, we can verify that the associated functional $\bphi_a\propto \epsilon_{ab} \vv{\dgap}^b$ is positive when acting on vectors with $\Delta>\dgap$. We have also checked that the result agrees with a linear programming computation (with bisection) to at least 15 digits (bisecting further than this takes a long time while we expect the flow answer to be accurate to much higher precision).

\begin{figure}
\begin{center}
\begin{tabular}{c}
\includegraphics[width=11.5cm]{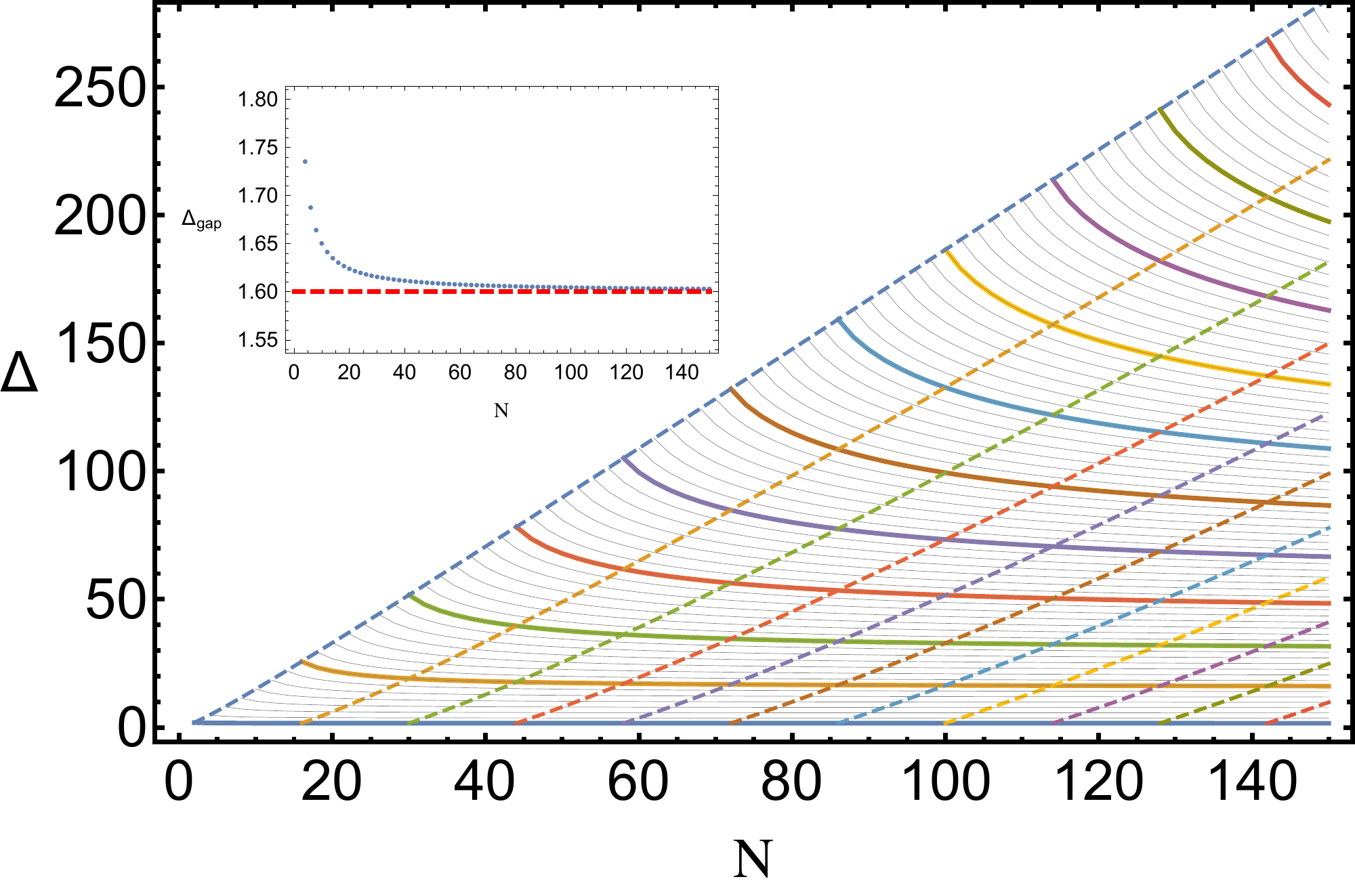}
\\
\includegraphics[width=11.9cm]{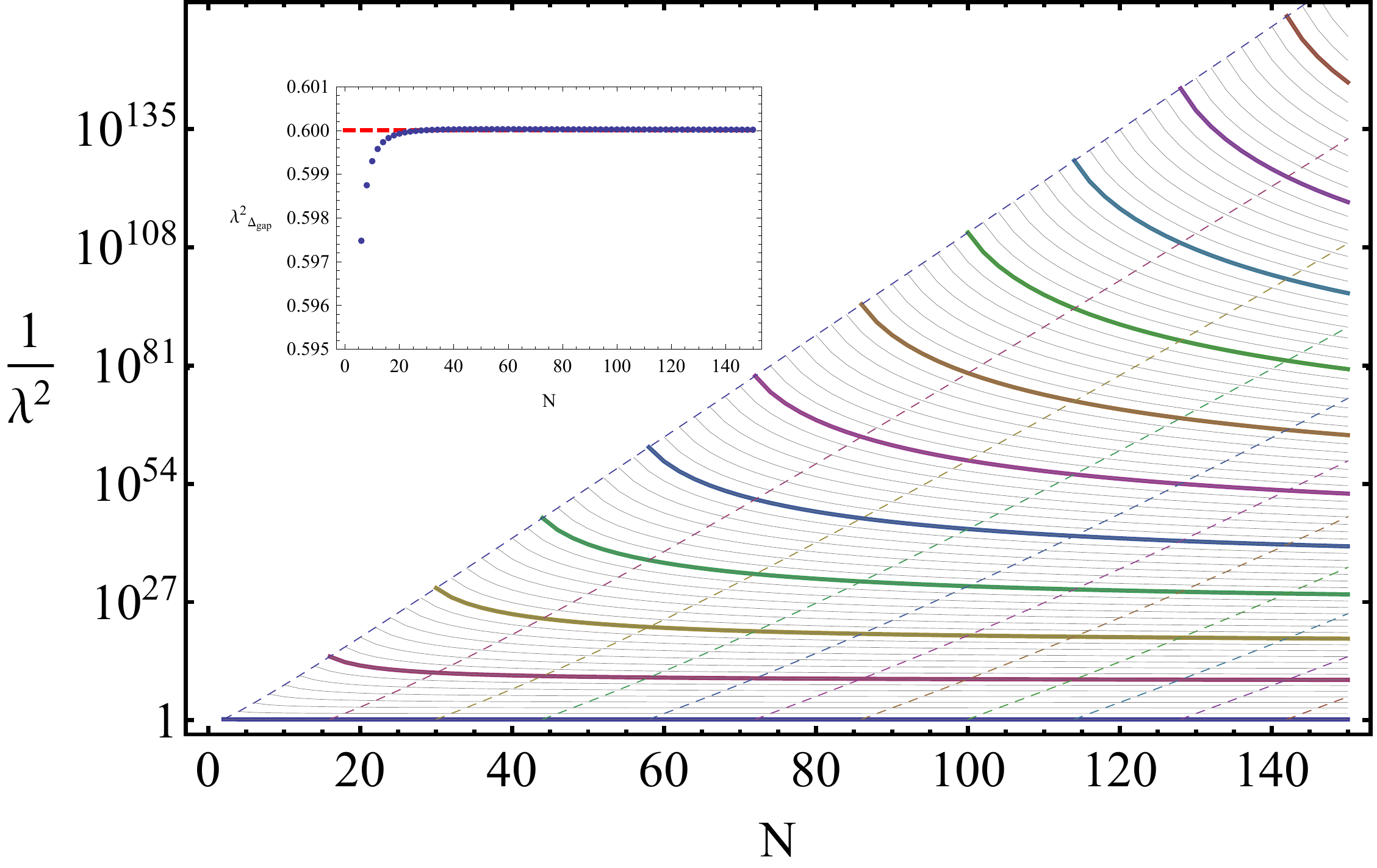}
\end{tabular}
\end{center}
\caption{Upgrading. Plots show the evolution of the spectrum as the number of crossing constraints $N$ is increased. At the top the conformal dimensions, and on the bottom the corresponding OPE coefficients. For each $N$, cutting the curves with a vertical line gives the spectrum at that $N$. For clarity a few chosen curves are highlighted in color. As $N$ increases new operators appear. Their dimension and OPE coefficient vary a lot in the beginning, but eventually stabilize. The diagonal dashed lines interpolate the successively {\em largest} dimension operators, and their OPE coefficients, as a function of $N$. The insets show the leading operator. Notice in particular the leading OPE coefficient converges very fast. Finally, for any $N$ the value $\dgap$ is a valid upper bound, which explains it's decrease with $N$.}
\label{fig:upgrading}
\end{figure}

The value $N=2$ is the smallest possible, and hence not particularly interesting. We would like to build on this solution to get higher values of $N$. In practice we have done this as follows. Firstly, we can restrict to even values of $N$. This is because every new operator in the extremal solution will be a double\footnote{This is an experimental observation. Also, for odd values, a  fixed vector appears at $\Delta=\infty$.}. For small enough values of $N$ we can get a new solution with $N+2$ components by simply guessing what the dimension of this operator will be, by taking it to be roughly 20\% larger than the current highest dimension. Error correction is then sufficiently powerful to find the exact solution. Once we have a few values of $N$, we can use those solutions as data for an extrapolation to the next value. Since our solutions are very accurate, this extrapolation can usually guess the correct spectrum to within $10^{-3}$, which is amply sufficient as an initial guess for error correction. One issue is that everytime we increase $N$ we need to add a new operator, whose dimension we also need to guess. The solution becomes clear by looking at the results, which are shown in figure \ref{fig:upgrading}. It shows the spectrum obtained by upgrading from $N=2$ to $N=150$ in steps of 2. Equivalently, the spectra contain operators from one to 75.
From the figure we see that the curves labeling highest dimension operators as a function of $N$ have a very simple behaviour, being nearly perfect straight lines (shown as dashed in the figure). It is these curves which we extrapolate, and in particular this allows us to very accurately guess the dimension of each new operator as it appears.

We emphasize that at no point here have we used linear or semidefinite programming methods. The overall run time on a single core processor was $\simeq$ 45 minutes\footnote{Computations were done in {\tt Julia} building on the {\tt JuliBootS} package \cite{Paulos2014a}, using 1000 bits of precision and conformal block representations including 200 poles.}. This of course gives us not only the final $N=150$ component results, but all the intermediate ones, which would need to be obtained separately with usual techniques. We have also checked that all solutions are extremal to very high accuracy. This means not only that we have an extremely accurate solution to crossing, but also a positive linear functional (with zeros at those vectors appearing in the solution). This functional guarantees that the value of $\dgap$ in each 
such solution is a valid upper bound for each value of $N$.

In particular, the determination of the values $\dgap$ for each $N$ is insanely accurate: better than one part in $10^{125}$ in our computation, the precision being limited only by our conformal block representations\footnote{The number quoted is a comparison between computations with 150 vs 200 poles \cite{Kos:2014bka}.}. Similar results are essentially impossible to achieve with ordinary techniques given the limitations in bisection. Indeed, we cannot even check our results at higher values of $N$ with linear or semidefinite programming, since bisection in $\dgap$ to reasonable accuracy takes an inordinate amount of time for higher values of $N$. However, for lower values of $N$ we did check that the (error-corrected) solution obtained from linear programming agrees with the one obtained by upgrading.

\begin{figure}
\begin{center}
\begin{tabular}{cc}
\includegraphics[width=7cm]{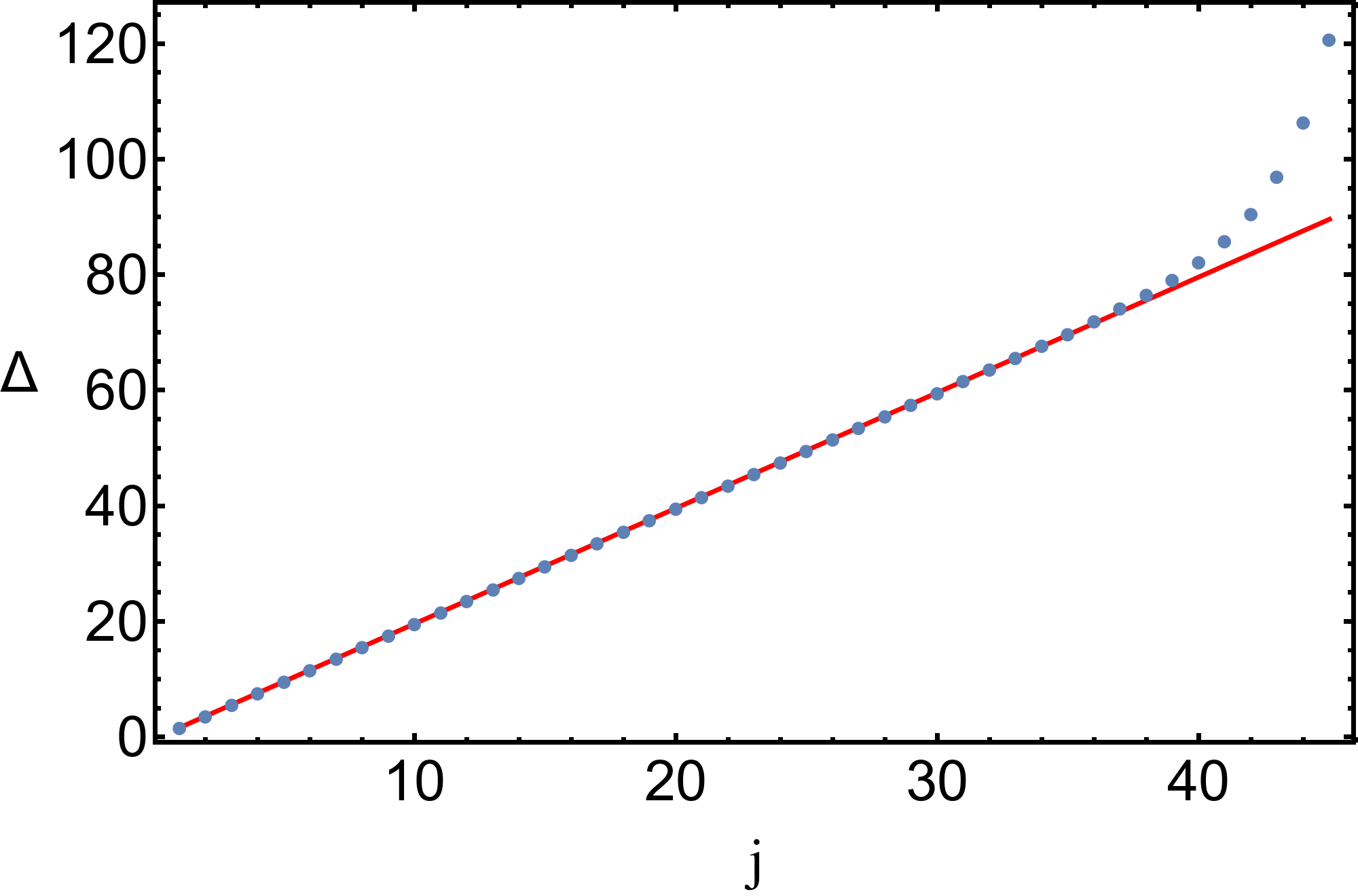}
&
\includegraphics[width=7.2cm]{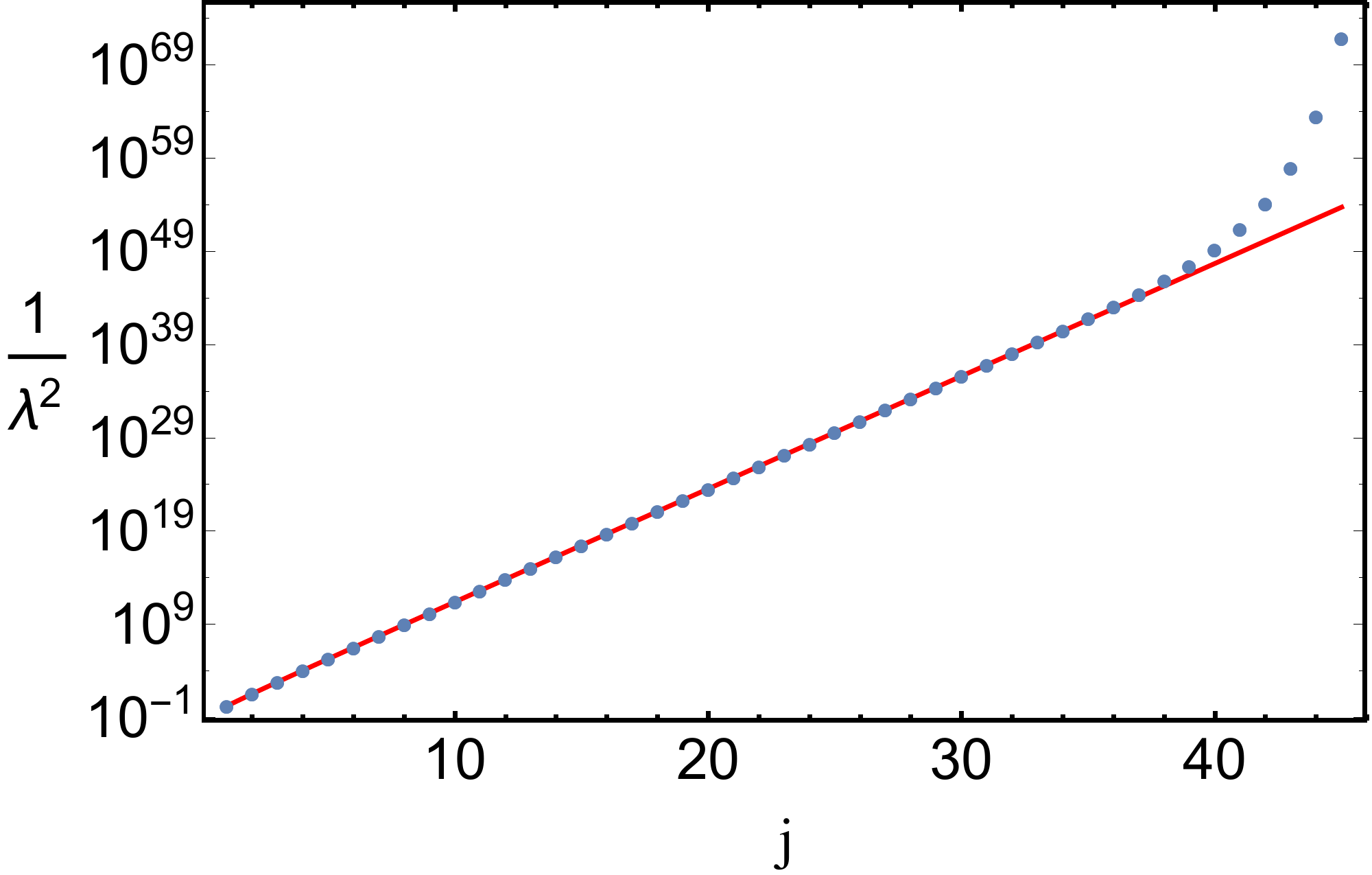}
\end{tabular}
\end{center}
\caption{Comparison between extrapolated vs exact spectra. The exact solution is the generalized free fermion with dimensions $\Delta(j)=1+2\Delta_\phi+2j$ (cf. \reef{eq:gff}), here evaluated at $\Delta_\phi=0.3$. The exact values lie on the solid red line whereas the extrapolated results are represented by the blue dots.}
\label{fig:compare}
\end{figure}

It is interesting to compare our results with those of the conjectured exact solution saturating the bound. Since in our computation we have obtained many highly precise spectra for several values of $N$, a natural thing to do is to extrapolate to $N=\infty$. We do this by taking the 45 lowest dimension operators, fitting their parameters with a high degree polynomial in $1/N$ and extrapolating to zero. The results are shown in figure \ref{fig:compare}. Overall we find excellent agreement, with relative errors ranging from $10^{-8}$ to $10^{-6}$ for the first 20 operators. What is important to take from this exercise is that it is the high accuracy of our results, guaranteed by our error-correction flows, which allows for such excellent extrapolations.
%
%

\subsection{Continuous flows}
\subsubsection{Gap maximization}
\begin{figure}
\begin{center}
\includegraphics[width=10cm]{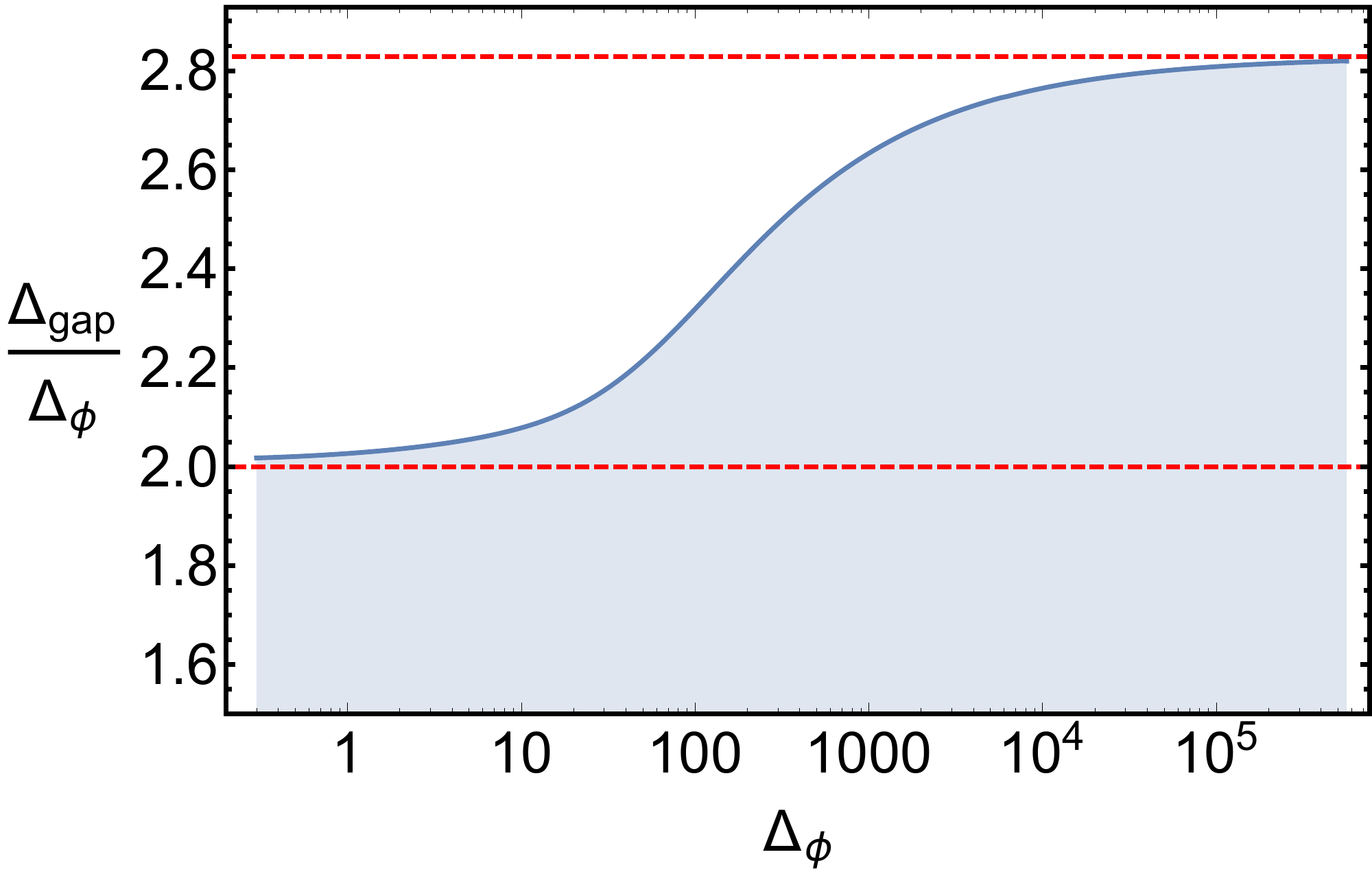}
\caption{Gap maximization with 100 components. The curve provides a valid upper bound on the dimension of $\phi^2$ in $D=1$ CFTs. The slope of the bound smoothly interpolates between $2$ and $2\sqrt{2}$. As the number of components increases, the transition region is pushed to higher values of $\Delta_\phi$.}
\label{fig:gapmaxbound}
\end{center}
\end{figure}

We shall now consider flows where we vary a continuous parameter. The simplest and most straightforward is a flow in the external dimension $\Delta_\phi$. In this case the perturbation of the crossing equations $\delta \mbf T$ is:
\bea
\delta \mbf T=\delta \Delta_\phi \frac{\partial}{\partial \Delta_\phi} \left(\sum_{i=1}^K a_i \vv i-\mbf T\right).
\eea
In evaluating this expression we are determining only the explicit variation with respect to $\Delta_\phi$. Solving the linearized flow equations gives us a new approximately extremal solution valid at $\Delta_\phi'=\Delta_\phi+\delta \Delta_{\phi}$, which is then error corrected. Starting from an upgraded solution with $N=100$ obtained following the methods of the previous section at $\Delta_\phi=0.3$, we can flow to other values. In this way we get a valid upper bound on the dimension $\dgap$ of the leading scalar as a function of $\Delta_\phi$, which is shown in figure \ref{fig:gapmaxbound}.

\begin{figure}
\begin{center}
\includegraphics[width=11.5cm]{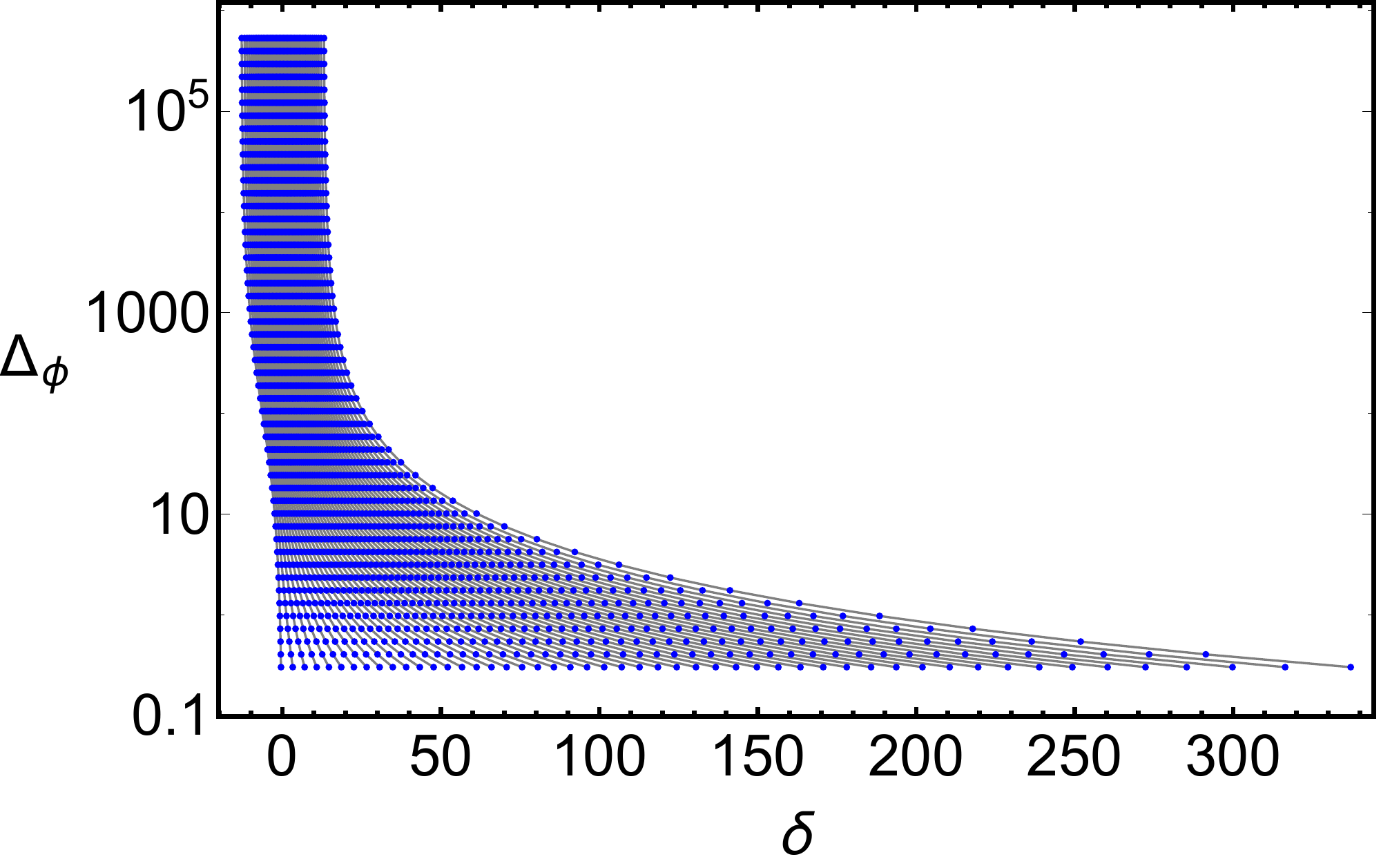}
\includegraphics[width=11.5cm]{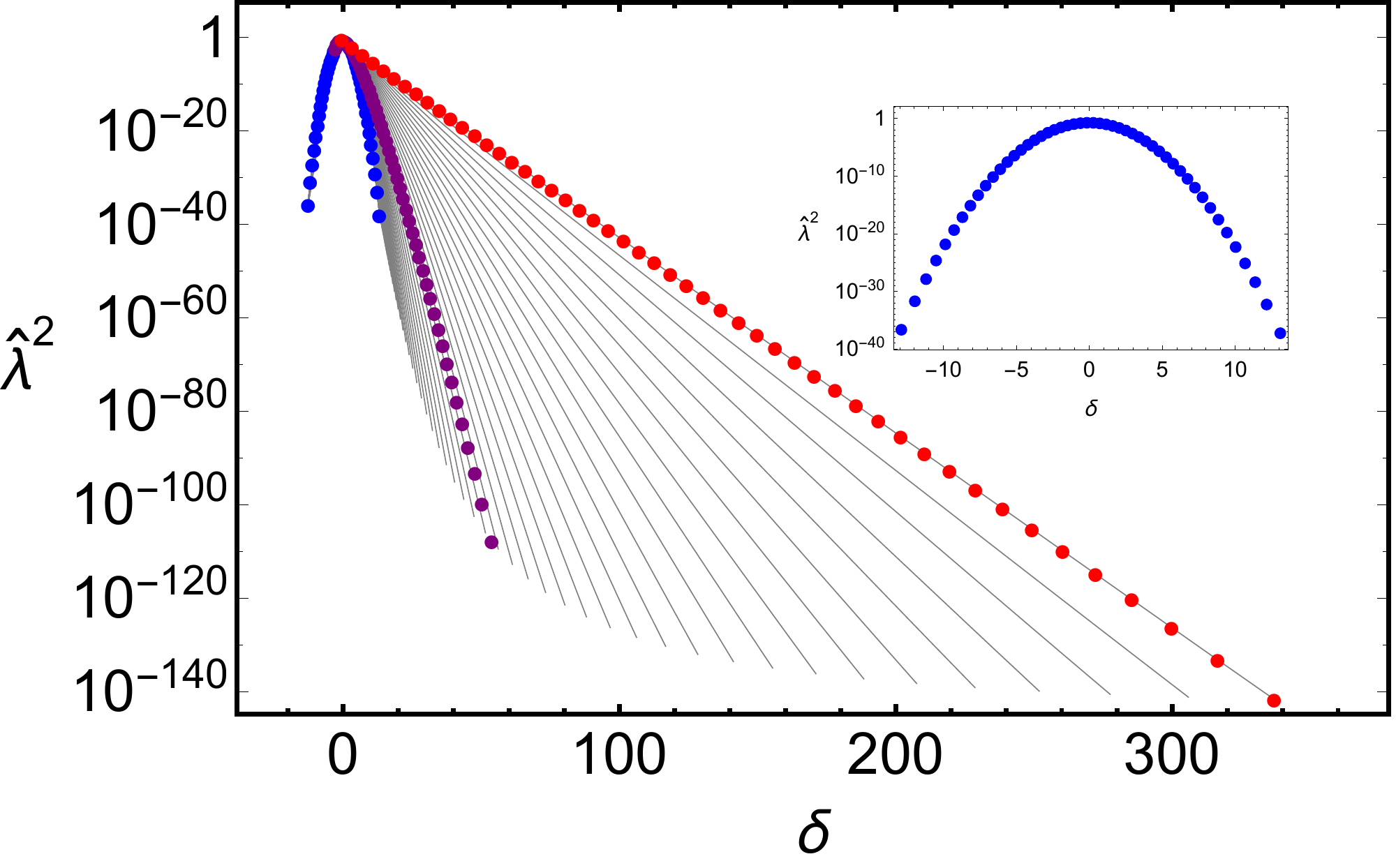} 
\caption{Gap maximization with 100 components. At the top, the flow of the spectrum of operator dimensions as $\Delta_\phi$ is increased. Dimensions of operators are shown in terms of $\delta\equiv (\Delta-2\sqrt{2}\Delta_\phi)/\sqrt{\Delta_\phi}$. Each horizontal slice is a spectrum at a given $\Delta_\phi$. For large values of $\Delta_\phi$ the dimensions of operators stabilize in a region with finite width in units of $\sqrt{\Delta_\phi}$ centered at $\delta=0$. On the bottom, the different curves show the OPE coefficients for various values of $\Delta_\phi$, moving to the left as it is increased. Each curve is actually made up of several points (some of which are shown explicitly) each corresponding to an operator in the spectrum with dimension $\delta$. The final configuration, at $\Delta_\phi\simeq 6\times 10^5$, is highlighted in blue and enlarged in the inset. It is very nearly gaussian with width $\simeq \sqrt{\Delta_\phi}$. The hats on OPE coefficients means we have written them in ``natural units''\cite{El-Showk2014a}, i.e. $\hat \lambda^2(\Delta)\equiv \lambda^2/(4\rho)^\Delta$, with $\rho=\frac{x}{(1+\sqrt{1-x})^2}|_{x=1/2}$.
}
\label{fig:gapmaxspec}
\end{center}
\end{figure}
Starting from $\Delta_\phi=0.3$ we have flowed all the way to $\Delta_\phi\simeq 10^6$. The bound curve shown in the figure contains around 200 points, all accurate to better than one part in $10^{100}$. Remarkably, the time to obtain each point ranges from one to three minutes on a single core, depending on the precision used. 
By contrast, standard bisection-based approaches using Linear Programming (at the same value of $N$) require over 2.5 hours to bisect a single point to one part in $10^6$.
We find experimentally that higher precision\footnote{Our computations range from 1000 to 1400 bits of precision. Such unusually high values are required in $D=1$, since for larger values of $N$ we have very high dimension operators in the spectrum.  We do not expect to require such high precision in $d > 1$.} is required for larger values of $\Delta_\phi$. Altogether these results show not only that extremal flows work in a continuous setting, but are orders of magnitude faster than traditional approaches. 

As a bonus, we can see that the slope of the bound seems to vary smoothly from about $2$ to $2\sqrt{2}$. In fact, the OPE coefficients seem fit a Gaussian curve centered around $2\sqrt{2}\, \Delta_\phi$ with a width of order $\Delta_\phi^{1/2}$, as shown in figure \ref{fig:gapmaxspec}. These results are in agreement with the general analysis of \cite{Kim:2015oca}. 

\subsubsection{OPE maximization}
This concludes our application of extremal flows to problem 1. Let us now turn to problem 2, which is OPE maximization. We would like to show that one can obtain an upper bound on an OPE coefficient as a function of $\Delta_\phi$ by a flow. There are different ways to approach the problem. We could obtain an initial extremal solution to OPE maximization by linear programming and then flow in $\Delta_\phi$. Here however we'll do something a lot more interesting: we will flow from problem 1 to problem 2. That is, we will start off with a solution which maximizes $\dgap$ and flow to a solution that maximizes an OPE coefficient.

To be definite, we will consider problem 2 (cf. \reef{eq:prob2}), maximizing the OPE coefficient $\lambda^2_{\Delta_\phi}$, i.e. with $\dels=\Delta_\phi$ and for various values of $\dgap$, with $\Delta_\phi$ fixed. Our goal is to find a solution to this problem starting from the corresponding solution to problem 1, which we found above. To see how this can be done, notice that there is a smooth family of solutions to OPE maximization labeled by $\dgap$. As we increase $\dgap$, the OPE coefficient decreases, until it eventually reaches zero. The corresponding value of $\dgap$ is precisely the solution to our first problem (assuming $\dgap > \dels$). Hence, we have only to reverse this logic: starting from this maximal case, we include an extra vector in the solution to crossing with dimension $\Delta_\phi$ and zero OPE coefficient. We then flow in $\dgap$, decreasing it little by little. As we do this the OPE coefficient increases. At every value of $\dgap$ the OPE coefficient is guaranteed to be the maximal one.

\begin{figure}
\begin{center}
\begin{tabular}{cc}
\includegraphics[width=7cm]{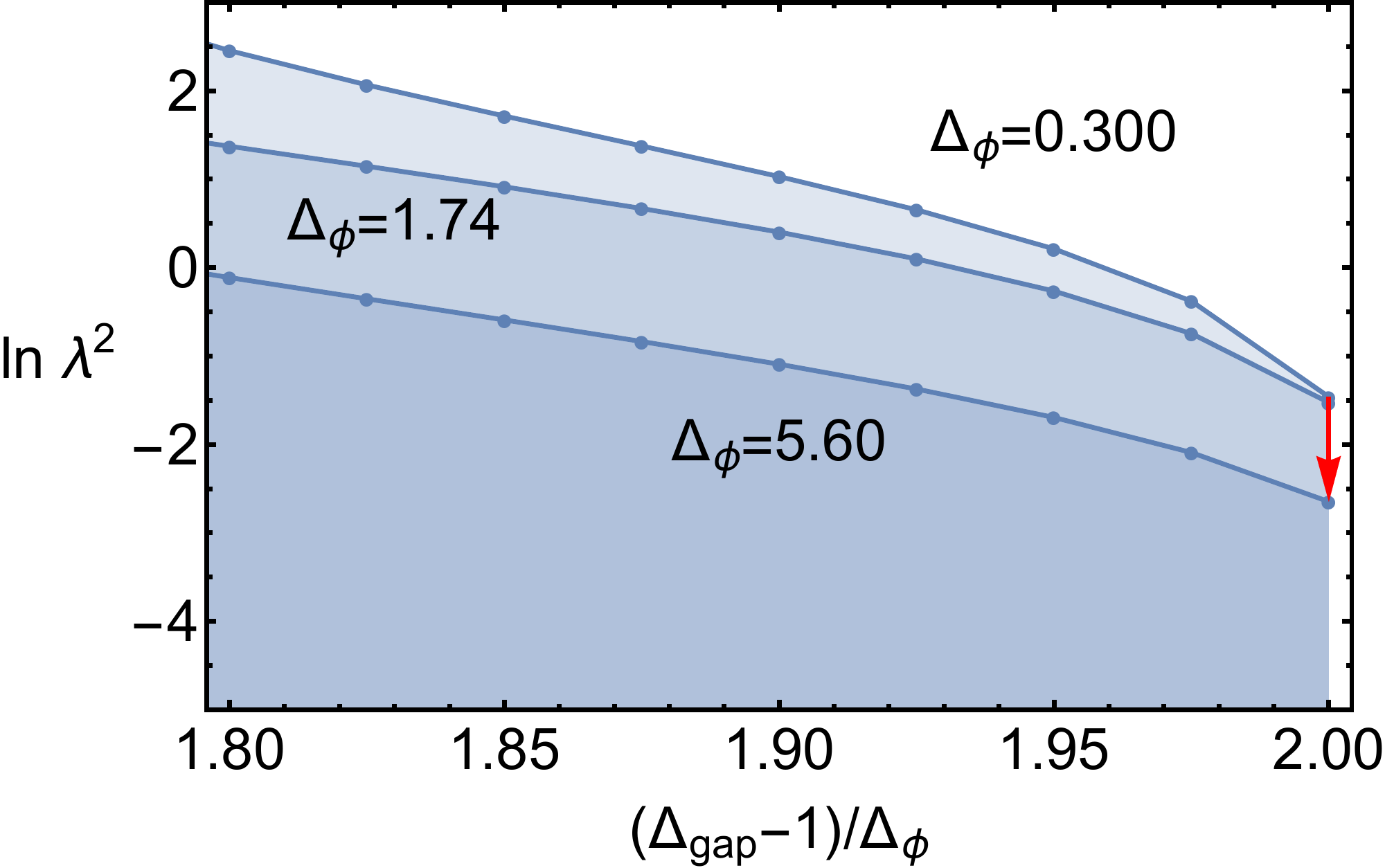}
&
\includegraphics[width=7cm]{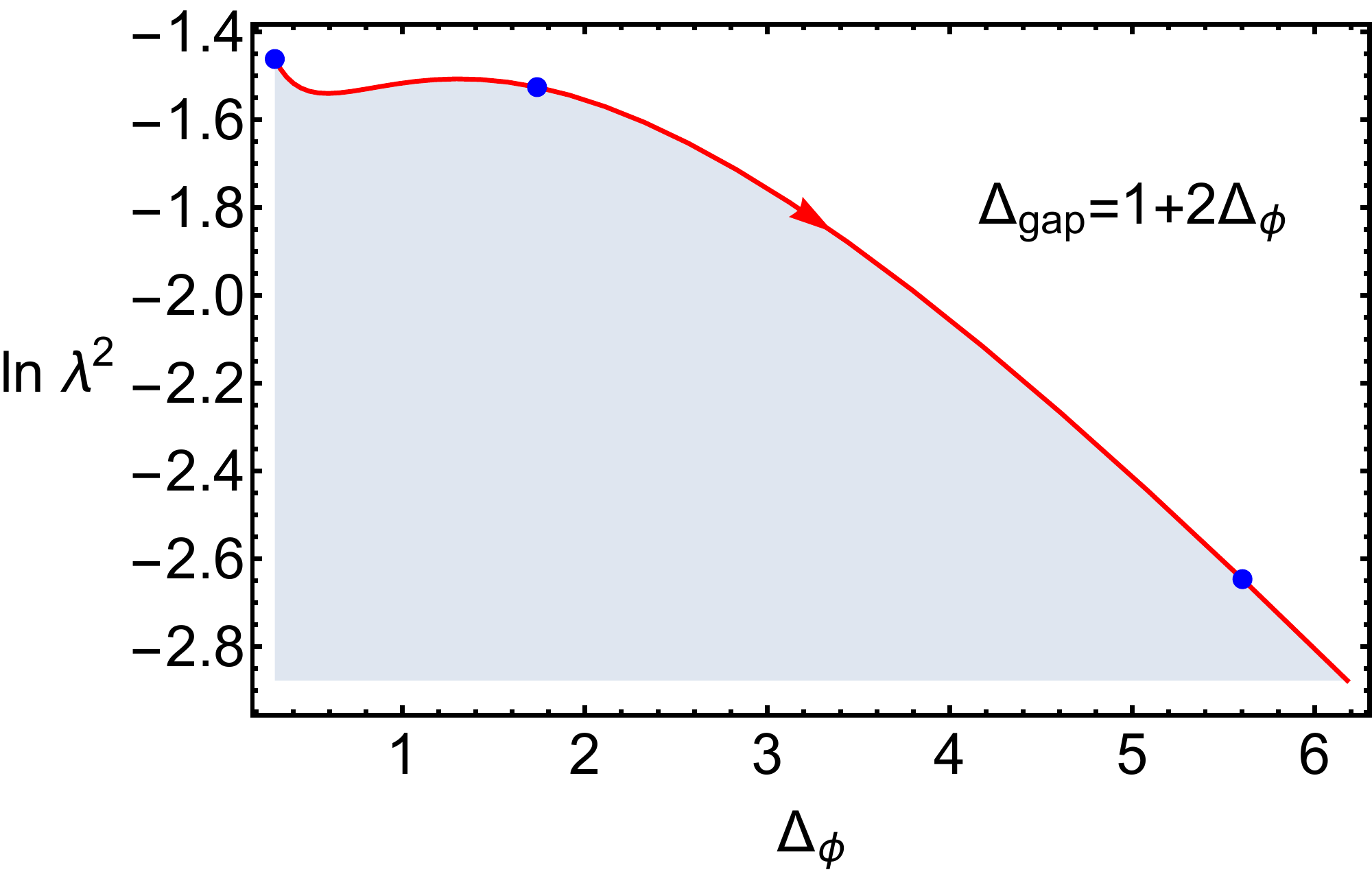}
\end{tabular}
\end{center}
\caption{OPE maximization with $N=100$ components. On the left, flows from gap maximization to OPE maximization. Given the OPE $\phi \times \phi=\phi+ \phi^2+\ldots$, with $\Delta_{\phi^2}\equiv \dgap$, we are placing an upper bound on $\lambda\equiv \lambda_{\phi \phi \phi}$. On the left, each curve corresponds to a different value for $\Delta_\phi$, and we vary the gap on the $x$-axis. On the right, we fix the gap instead to $(\dgap-1)/\Delta_\phi=2$ and flow in $\Delta_\phi$. Hence the red curve on the left plot should match the one on the right. In particular, the blue dots correspond exactly to the intersection of the three curves on the left with the vertical line at $\dgap=1+2\Delta_\phi$.}
\label{fig:opemax}
\end{figure}

Doing this for several values of $\Delta_\phi$ leads to a set of upper bounds on $\lambda^2_{\Delta_\phi}$ as a function of the gap. Some of these are shown in figure \ref{fig:opemax}. As a consistency check, these different curves should be connected by flowing in $\Delta_\phi$ keeping the gap parameter fixed. This is also shown in the same figure for a particular value of the gap. The two approaches are of course compatible. 

The run times are similar to the previous section. In particular, since here we have not considered large values of $\Delta_\phi$, we can work with smaller values of precision. Each point takes then between 30 seconds to a minute. At every point we can check extremality both by verifying that crossing is satisfied and that the associated extremal functional is positive everywhere. In all cases this can be verified to extremely high accuracy. This shows that extremal flows can be used, and are very efficient, in the context of problem 2, i.e. OPE maximization.

\section{Discussion}\label{sec:disc}
\subsection{Non-unitary flows and the method of determinants}\label{sec:det}
We would now like to comment on the relation between our formalism and the method of determinants, introduced by Gliozzi in \cite{Gliozzi2013} and developed further in \cite{Gliozzi2015,Gliozzi2014}. The starting point of this method is the set of truncated crossing equations introduced in sections~\ref{sec:efm} and~\ref{sec:ext}:
\bea
\sum_i a_i   \mbf v_i=  \mbf T. \label{eq:crossf}
\eea
In the above, unitarity would usually restrict the set over which $i$ can range, and also sets $a_i\geq 0$. We would like to see what we can say about solutions to these equations, without necessarily assuming unitarity. 

For definiteness we take all vectors to have $N$ components. Without assuming unitarity, the equations \label{eq:crossf} have an infinite set of solutions. To see this it is sufficient to take any generic subset of $N$ vectors, which will inevitably form a basis, and since the equations are linear the solution is trivial. Clearly the problem is overdetermined, and to make progress we need to make some restrictions. In particular we may well wonder whether the equations would still have solutions if we only allow a number of vectors $K$  which is strictly smaller than $N$. CFTs which admit such truncated solutions have been called ``truncable'' \cite{Gliozzi2013}. As will become clear, truncability is nothing but a special case of extremality.

So, let us take some set of $K$ vectors and ask if a solution exists. One idea is that if a set of $N$ linear equations has a solution involving $K$ vectors, then putting these vectors together with $\mbf T$ into a matrix, all $(K+1)\times (K+1)$ subdeterminants should vanish. This is what gives the determinant method its name. From our point of view such a procedure is not very satisfactory: the number of possible subdeterminants rapidly grows with $N$ and $K$, but not all of them are independent. Furthermore a trivial solution is always to take two identical, or proportional, vectors. There is however a simpler way. Introduce $N-K$ auxiliary vectors $ \mbf w_j$, for instance some of the columns of the $N\times N$ identity matrix. Then it is straightforward to solve%
\bea
\sum_{i=1}^K a_{i}   \mbf v_{i}+\sum_{j=1}^{N-K} \mu_{j}  \mbf w_{j}= \mbf T.
\eea
Clearly then the conditions one should require are that $\mu_j=0$ for all $j$, which imposes only $N-K$ constraints overall. If we wish, we may also rephrase these conditions in terms of determinants,
\bea
\langle \vv 1\ldots \vv K\,  \mbf w_1 \ldots \widehat{ \mbf w_j} \ldots  \mbf w_{N-K} \,  \mbf T\rangle =0, \qquad j=1,\ldots, N-K.
\eea
Of course one should not forget the overall non-singularity condition,
\bea
\langle \vv 1\ldots \vv K\,  \mbf w_1 \ldots  \mbf w_{N-K} \rangle \neq 0,
\eea
which rules out degenerate solutions.

Proceeding, let us suppose we have found some solution to these constraints. We must now determine how unique this solution is. By this we mean whether it is possible to smoothly deform the solution to some new set of vectors and OPE coefficients satisfying the same equations, and if so, what is the dimensionality of this solution space. This is easy to determine. There are $K$ coefficients $a_i$, and let us assume that the label $i$ includes $C$ continuous parameters -- namely the conformal dimension, and $C-1$ angular variables. Then the dimensionality of a connected component of the space of solutions is simply  the number of degrees of freedom minus the $N$ constraints \reef{eq:crossf}, that is $(1+C)K-N$. Hence we expect a unique solution only if $K=N/(1+C)$, and if the solution space is connected. These simple observations explain the issues found in \cite{Gliozzi2013,Gliozzi2014,Gliozzi2015}. There typically $C=1, K=N-1$, and so without extra restrictions on the degrees of freedom (such as fixing some conformal dimensions), one rapidly runs into non-uniqueness even for small values of $N$. 

From our perspective, it is clear what is going on. Extremal solutions are singled out by the full set of extremality conditions \reef{eq:extreme}, which includes not only crossing but also the saturation and especially the tangency conditions. Hence, the determinant method can only work in the very special case where the crossing solution includes no singles in the spectrum, since in this case the tangency conditions can be satisfied trivially. In particular this requires 
$K\simeq N/2$ and not $K\simeq N$ as is usually used. Some leeway can be gained by inputing some information about the theory that one wants to study, such as fixing the presence of certain operators in the spectrum. This reduces the number of degrees of freedom by hand and allows one to push the method to higher values of $N$. Eventually however the information one has available quickly runs out and one is left with indeterminacies.

Our own method suggests a way out: one should simply add the tangency conditions to the crossing equations. Naively there may seem to be no reason for doing so. After all the tangency conditions arose as a consequence of maximization together with positivity. However, a different point of view is to see these two requirements as a scaffold that allows us to arrive at the tangency conditions, which in the end are simply the statement of certain linear dependencies amongst vectors and their gradients. 

This perspective has its own problems but it seems natural and relatively promising. When doing a flow, it is  possible that an OPE coefficient becomes zero. Demanding unitarity this presents some kind of singularity in the flow, as shall be discussed in section \ref{sec:sings}, which forces us to move to a new branch of solutions. However, one could simply continue flowing along the same direction. In this way one arrives at extremal solutions, in the sense of satisfying all the extremal equations, but with an OPE coefficient that is now negative. It seems natural to think that such non-unitary solutions should be singled out as special.

\subsubsection{The non-unitary GFF in $D=1$}
As a concrete test of our proposal, we will bootstrap, for the first time, the non-unitary generalized free fermion in $D=1$. This is simply the analytic continuation of the solution shown before to negative values of $\Delta_\phi$. Besides this fact, non-unitarity also shows in the fact that the leading OPE coefficient in the four point function, $\lambda_{j=0}^2$ is negative, as follows from \reef{eq:gff}. Hence, such solutions are not accessible with linear or semidefinite programming methods. Our challenge is to obtain it by using flows.

\begin{figure}
\begin{center}
\includegraphics[width=11.5 cm]{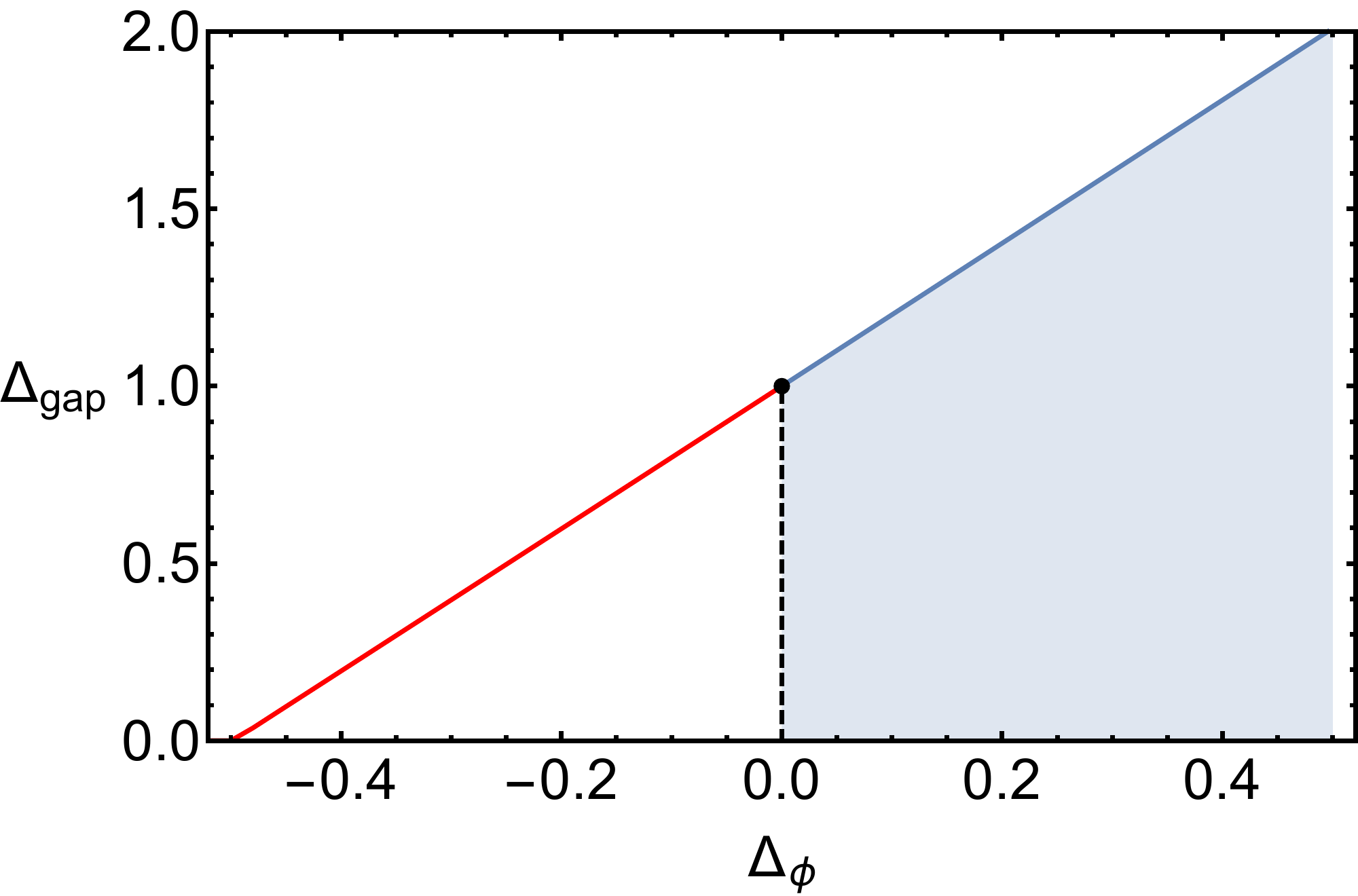}
\caption{Flowing into a non-unitary region with $N=100$ components. Below $\Delta_\phi=0$ the extremal solution develops a negative OPE coefficient. However, there is still an associated positive linear functional. The functional does set a bound on possible unitary solutions in this region, but this bound may not be optimal.}
\label{fig:nonunitflow}
\end{center}
\end{figure}

\begin{figure}
\begin{center}
\begin{tabular}{cc}
\includegraphics[width=7cm]{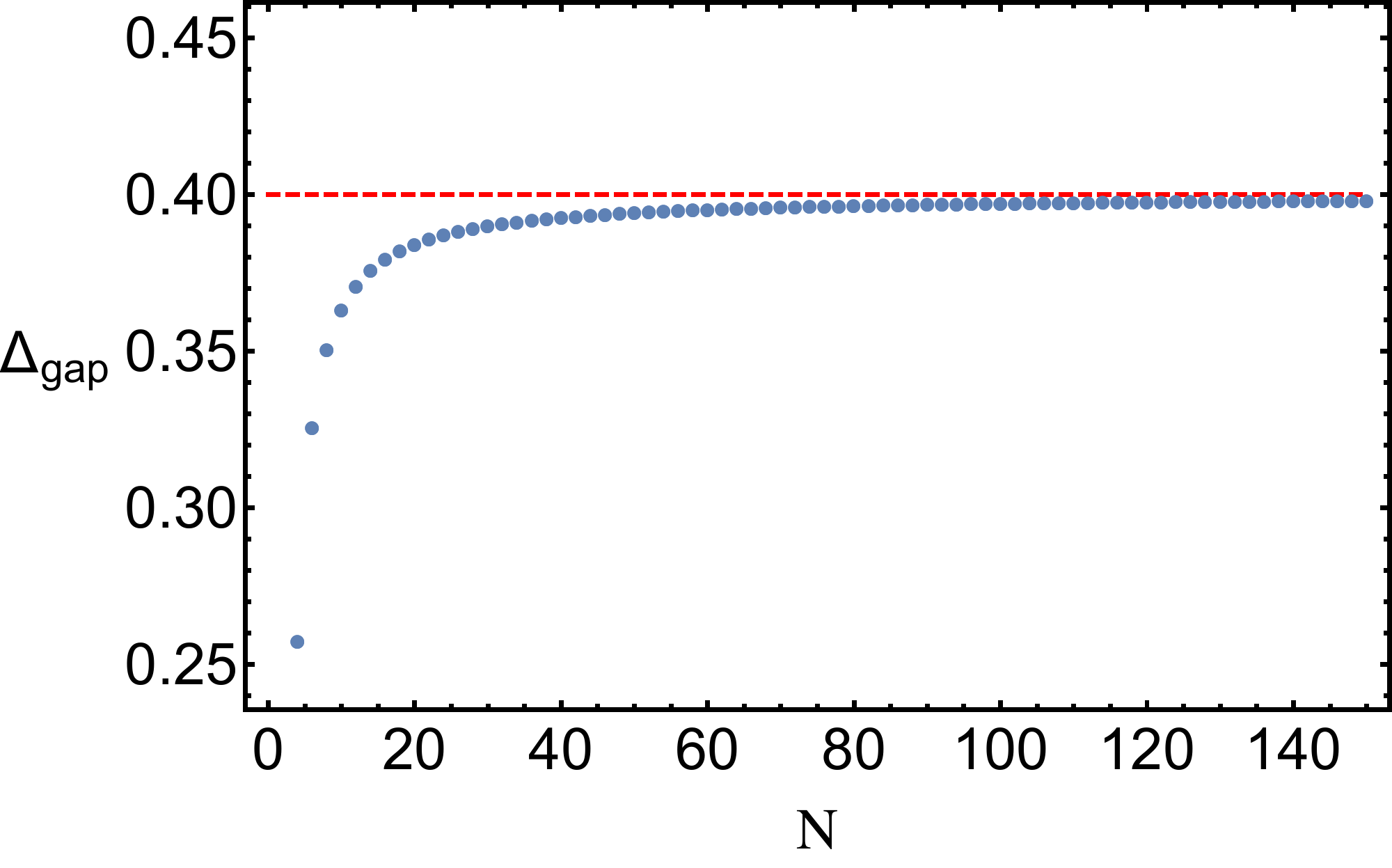}
&
\includegraphics[width=7.45cm]{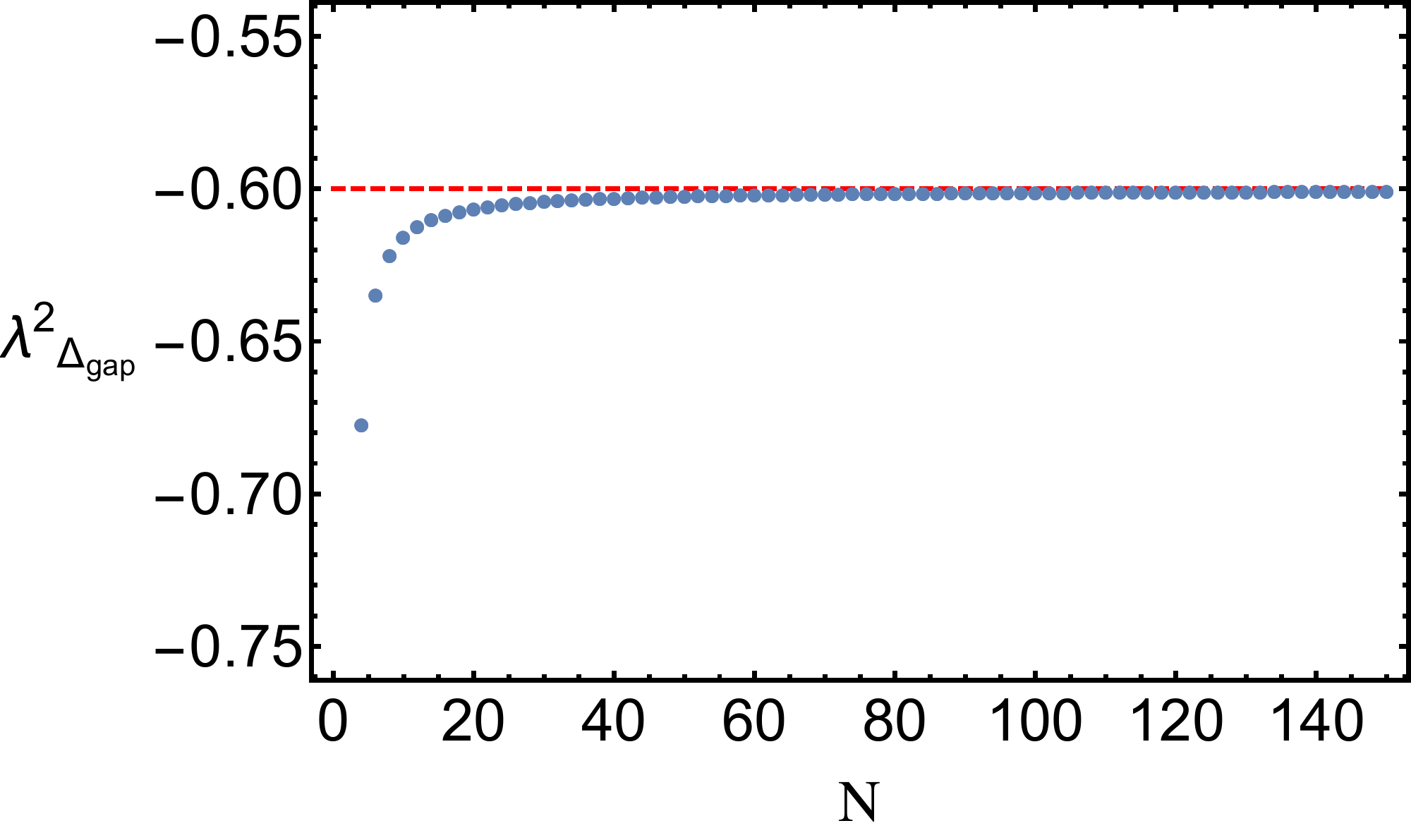}
\end{tabular}
\end{center}
\caption{Upgrading at a non-unitary point, with $\Delta_\phi=-0.3$. On the left, evolution of $\dgap$ as we increase the number of crosing constraints. The values seem to converge to the correct value $1+2\Delta_\phi=0.4$. Unlike the usual unitary bootstrap, the curve here does not have a meaning of a bound. Accordingly the value $\dgap$ does not need to decrease as we add more constraints, and in fact here it does the opposite. On the right, the leading, negative, OPE coefficient squared, compared with the exact value $\lambda^2_{j=0}=2\Delta_\phi=-0.6$.}
\label{fig:boundsnonunit}
\end{figure}

\begin{figure}
\begin{center}
\begin{tabular}{cc}
\includegraphics[width=7cm]{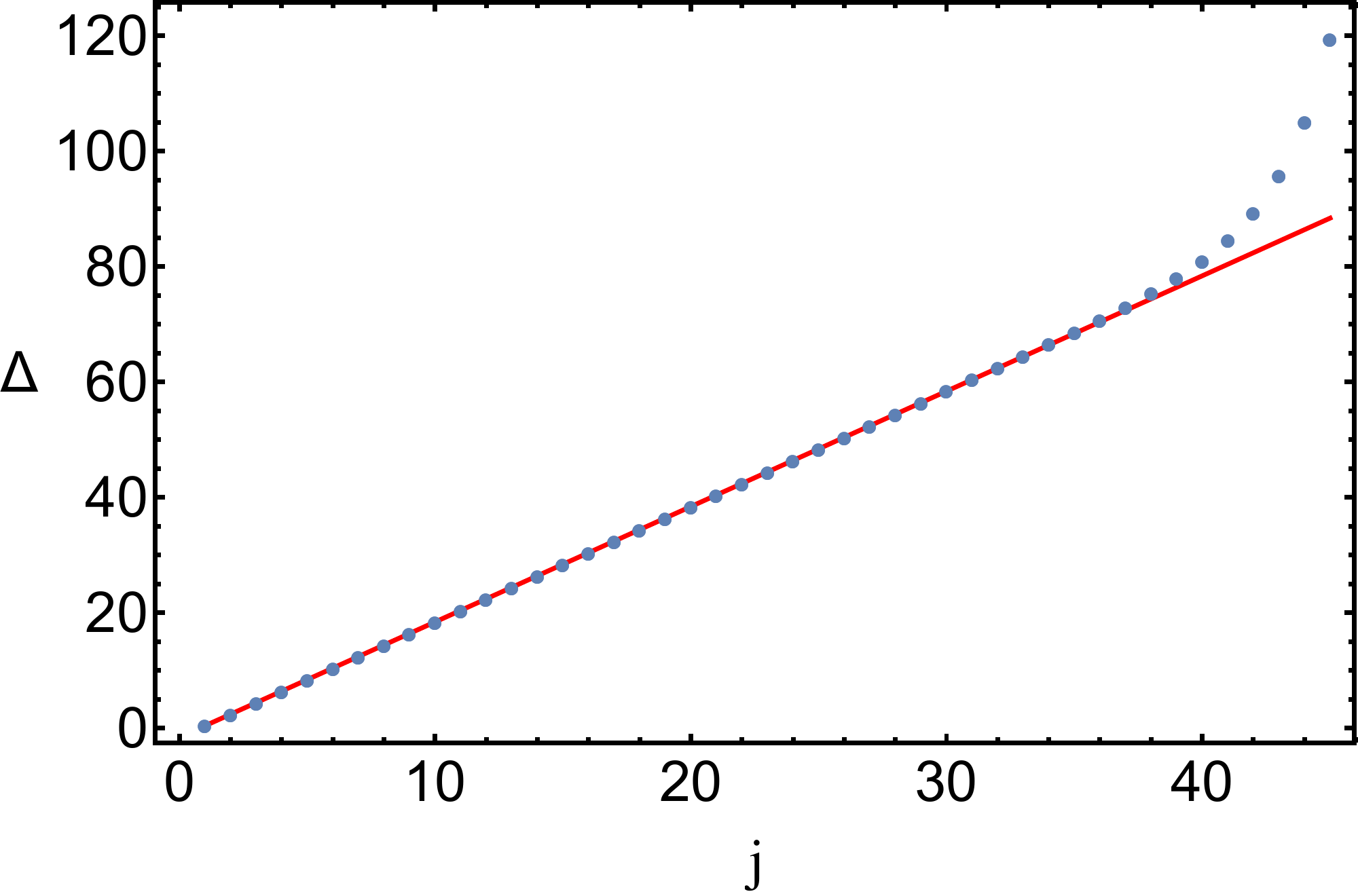}
&
\includegraphics[width=7.2cm]{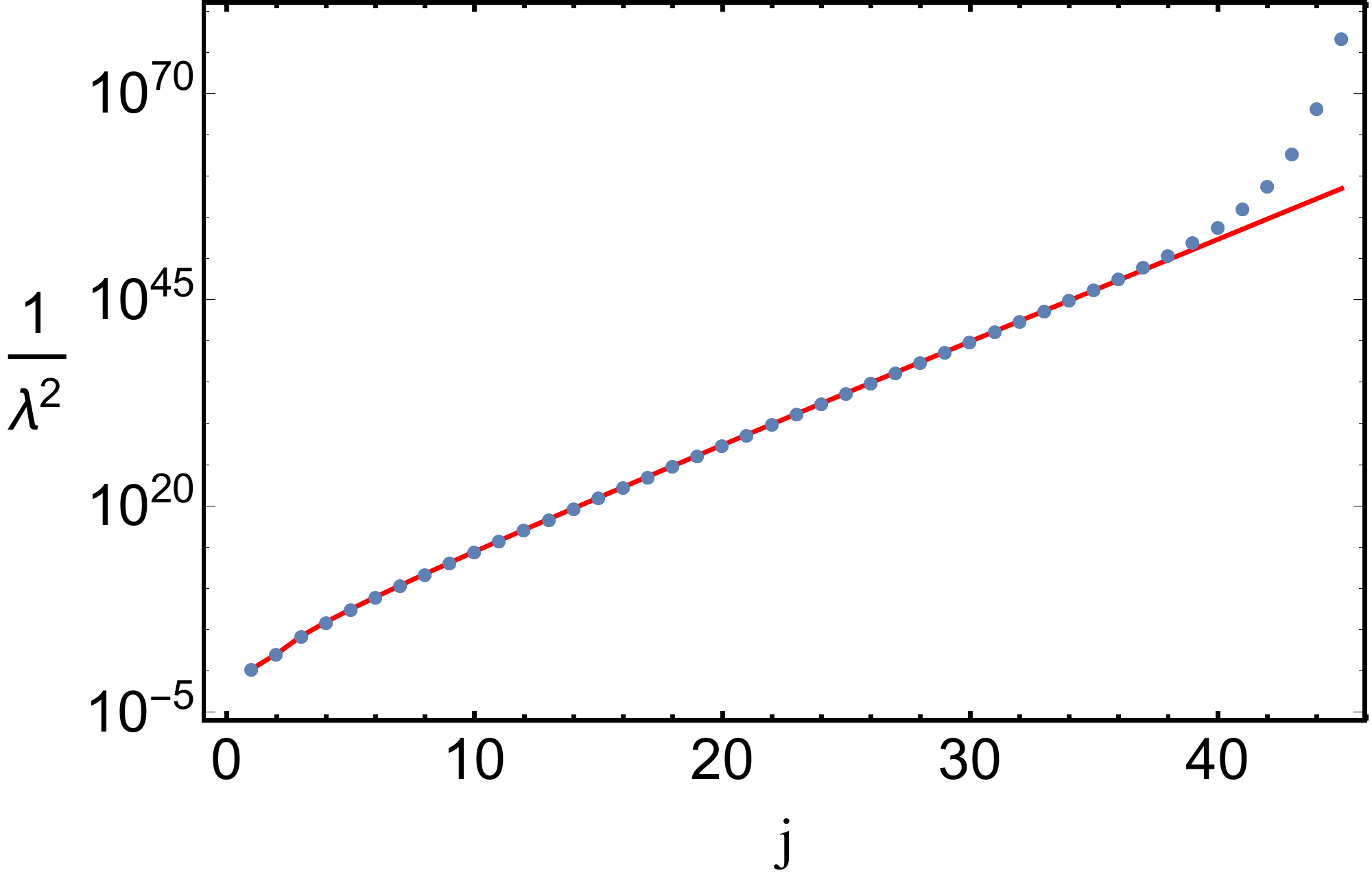}
\end{tabular}
\end{center}
\caption{Upgrading at a non-unitary point: Comparison between the extrapolated spectrum (blue dots) and the exact generalized free fermion (red line). Operators are labeled by an integer $j$, with $\Delta_j=1+2\Delta_\phi+2j$. The first 20 operators are correct to better than a part in $10^6$. Note that on the right $\lambda^2<0$ for $j=1$ (we are showing it's absolute value).}
\label{fig:extrapnonunit}
\end{figure}

We can do this in two ways, following the same methods as in section \ref{sec:app1d}. Firstly, we can simply extend our gap maximization results to negative values of $\Delta_\phi$, by flowing to this region. Of course, the minute we reach negative values, our results no longer have an interpretation as bounds. Rather, as we've discussed above, they should be thought of as capturing particularly simple, sparse solutions to the crossing equations.  This is straightforward to do and leads to a curve following very closely the expected line $\dgap=1+2\Delta_\phi$, terminating at $\Delta_\phi=-1/2$, as shown in figure \ref{fig:nonunitflow}. 
The second method is to upgrade directly a solution at some fixed $\Delta_\phi$. This proceeds exactly in the same way as in the unitary case, and the results are shown in figures \ref{fig:boundsnonunit}, \ref{fig:extrapnonunit}. The agreement with the exact non-unitary solution is as impressive as in the unitary case, and we can easily get around 20 operators correct to within one part in a million.

These results provide a first test of our approach to the non-unitary bootstrap. The same results could be in principle obtained from the determinant method by setting the number of operators equal to half the number of constraints. This is because here there are no single operators, and hence the tangency conditions are unnecessary. The extremal functional is constructed from a solution to crossing as in \reef{eq:func}, with $n_s=0$. Curiously, these extremal functionals are positive everywhere above $\dgap$ for each $N$, as we have checked. However, since we are allowing for negative OPE coefficients squared, this does not imply a bound of any sort.  

\subsection{Convergence}

We have seen that both in the usual, unitary bootstrap as well as in the determinant method, one has access to approximate solutions to crossing symmetry, that satisfy some truncated set of constraints. Here we would like to understand how well one can expect such solutions to approximate the actual spectra of CFTs. Our remarks will be somewhat heuristic, but hopefully they will inspire a more precise analysis in the future.

Let us start off with an exact solution to crossing constraints and truncate these down to $N$ components, which we choose to be derivatives with respect to cross-ratios. The crossing equations can be written exactly as
\bea
\sum_{i=1}^{K_{\dels}} a_i \vv i
=
\mbf T-\boldsymbol{\epsilon} \label{crossingeq}
\eea
and $ \mbf v_1, \ldots  \mbf v_{K_{\dels}}$ is the set of vectors with dimension smaller than some $\dels$. That is, $K_{\dels}$ counts the number of operators with $\Delta<\dels$. The error term is
\bea
\begin{array}{c}
\partial\\
\vdots \\
\partial^{N}
\end{array}
\left(
\begin{array}{c}
\epsilon_1 \\
\vdots \\
\epsilon_{N}
\end{array}
\right)
\equiv \boldsymbol{\epsilon} =\sum_{\Delta>\dels} a_i  \mbf v_i
\eea
 As written, the equations are exact, and hold for any truncation and any $\dels$ of any solution to crossing symmetry. In the bootstrap we obtain extremal solutions to crossing symmetry where at the very most one gets as many vectors as components; usually less. Hence we set $K_{\dels} \to K\leq N$ in numerical applications. Of course we do not have access to the error $\beps$. Hence, we are left with computing solutions to crossing by setting $\beps\to 0$ and hoping the error is not too large, but this will necessarily depend on the choices of $K,N$.

In the absence of derivatives it has been shown \cite{Pappadopulo2012} (see also \cite{Rychkov2016,Kim:2015oca}) that the error decays exponentially:
\bea
\epsilon_0\lesssim O((\dels)^{2\Delta_\phi} C^{\dels}), \qquad C<1.
\eea
The constant $C$ depends on the specific value of the cross-ratio\footnote{This cross-ratio is given in terms of $u,v$ in the usual way, $u=z\bar z$, $v=(1-z)(1-\bar z)$.} $z$ at which we evaluate the crossing relation. Since taking derivatives brings down factors of $\Delta$ from the $z^\Delta$ factor in a conformal block it is natural to expect that
\bea
\epsilon_p\lesssim O((\dels)^{2\Delta_\phi+p} C^{\dels}), \quad C<1
\eea
and indeed we show this in appendix \ref{sec:errder}.

We now ask the question: for fixed $N$, how large must we take $\dels$ such that all the components of $\beps$ are small?. We expect that it is sufficient to demand that $\epsilon_{N}\ll 1$. Given the error estimate above we must take
\bea
\dels \simeq \frac{N}{-\log(C)}. \label{eq:errdels}
\eea
Suppose now that $K_{\dels}$ grows as $(\dels)^\alpha$. This means that with $K$ vectors we can go up to a cutoff $\dels\simeq K^{1/\alpha}$. Then given \reef{eq:errdels}, with such a cutoff we expect that the error will be small for a number of components $N_{\mbox{\tiny eff}}$ roughly given by
\bea
N_{\mbox{\tiny eff}}\simeq \dels \simeq K^{1/\alpha}
\eea
In numerical applications, the best case scenario is $K=N$. The conclusion is then that the number of components for which the error is guaranteed to be small, $N_{\mbox{\tiny eff}}$, is related to the size of the truncation $N$ by:
\bea
N_{\mbox{\tiny eff}} \simeq N^{1/\alpha}
\eea
In other words, given an exact solution to crossing involving $N$ vectors, the error made by setting $\vec{\epsilon}\to 0$ in equation \ref{crossingeq} is guaranteed to be small for approximately the first $N^{1/\alpha}$ components. We can imagine starting from this equation and slowly sending $\beps\to 0$. This sets up a flow which can be solved using our formalism. We see that typically most of the vectors, especially those with high dimensions will move around a lot when we do this. Only a fraction is expected to remain approximately unmodified. In particular, increasing $N$ leads to a higher and higher number of operators that are distant from the correct solution. In principle we must then work hard to obtain a small fraction of correct operators. For example, in the generalized free field we have $\alpha=2$, so the number of correct operators is expected to grow like the square root of the size of the truncation. In light of the analytic bootstrap results of \cite{Komargodski2013,Fitzpatrick2013}, in {\em any} CFT in $d>2$ (apart from free field theory) $\alpha$ is at least 2.

In reality, this may even be too optimistic. The actual number of CFT primary operators grows exponentially fast with $\dels$. Indeed from the form of the high temperature limit of the entropy\footnote{In a CFT containing a stress tensor, the entropy on plane goes as $S\propto T^{d-1}$ and the free energy like $E\propto T^d$.}, we find the total number of states
\bea
N(\dels)\propto \exp(\gamma (\dels)^{1-1/d})
\eea
This huge number of states cannot be accounted by descendants of primaries which grow only like a power of $\dels$ (specifically $\dels^{d-1}$). Bootstrapping a correlation function for such a CFT is probably hopeless in the long run, unless some miracles happen. 

What sort of miracles? Well, firstly the number of primaries that actually appear in a specific correlator can be much smaller than exponential. For free theory, $N(\dels)\propto \dels$ and for the generalized free field $N(\dels)\propto (\dels)^2$. For the 2d Ising model, something else happens. Although the number of primaries increases exponentially, the number of {\em distinct} operator dimensions actually increases like $N^2$ (actually like $N$, but we distinguish operators with different spin). This is due to the magic of Virasoro symmetry that causes lots of operators to have coincident dimensions. The decoupling of null states is not the reason for this, rather it is the fact that the correlation function only contains two Virasoro primaries, and everything else has dimensions shifted by integers from this. In this sense, all minimal models have ``minimal'' correlators. and they have a chance to be bootstrappable.

This analysis makes it all the more remarkable that very accurate results have been obtained for the low-lying spectrum of a variety of CFTs, like the 3d Ising model. The most prosaic (and likely) explanation is that such theories have symmetries (like $Z_2$ for the Ising model) and perhaps other decoupling conditions which cut down the number of operators appearing in a given correlation function, and this is sufficient to pin down the properties of low-lying operators. 

\subsection{Singularities}\label{sec:sings}
We now turn to an important point which is the question of singularities in the flows. This issue was briefly mentioned in section \ref{sec:det}. Here we will undertake a more general discussion, which will hold in applications of the method to higher dimensions \cite{futurework}.

Firstly, a question of nomenclature. By singularity we simply mean a non-analyticity along a {\em unitary} flow, that is, a flow where positivity is maintained throughout. Such features occur  precisely because positivity conditions are non-analytic. The most notable (and useful) examples of singularities are kinks \cite{Rychkov:2009ij}, but even discontinuities can occur \cite{ElShowk:2012ht}. We say useful because such singularities seem to signal the presence of interesting theories and provide a useful criteria for determining their properties. While kinks have generally been the bootstrap stars, we believe there are several other kinds of more elusive singularities which have so far escaped our attention, and which could also point to interesting theories. Let us discuss the most interesting cases, leaving a detailed analysis for future work.

\subsubsection*{1. One operator more}

\begin{figure}
\begin{center}
\includegraphics[width=12cm]{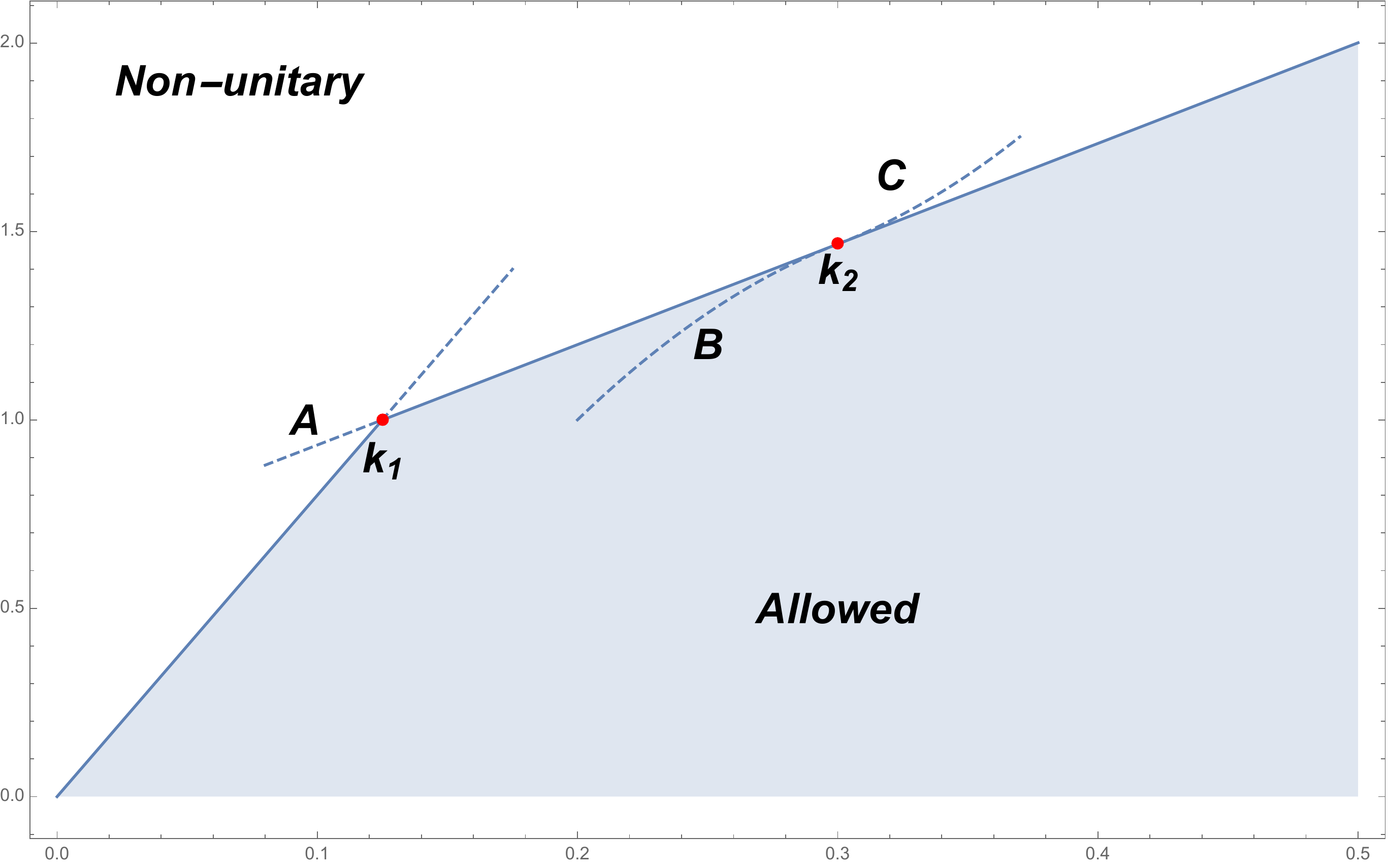}
\caption{Singularities and flows (schematic). At an ordinary kink, such as $k_1$ it is possible to flow into a non-unitary region, say along $A$, where some OPE coefficient becomes negative. However, not all decouplings of operators signal a visible kink. At when we approach $k_2$ from the left, an operator decouples. By allowing its OPE to become negative we can flow along $C$. Conversely, coming from the right, the functional develops a new zero at $k_2$. If we do not input this new vector into the solution, the latter won't be extremal anymore, and the flow will take us along $B$. }\label{fig:sings}
\end{center}
\end{figure}

When solving the flow, it is not necessarily guaranteed that we are finding the extremal solution. This is best seen from the functional perspective. Indeed, suppose at some point in the flow the functional acquires a new zero, i.e. some vector $\mbf v^*$ not present in the solution to crossing appears for which $\bphi\cdot \mbf v^*=0$. At this point, if we continue the flow in the same way, the functional will develop a negative region, i.e. we will have $\bphi\cdot \mbf v^*<0$. Note that we will still have a good solution to crossing as well as a functional satisfying all the tangency conditions. However, it will not be the extremal one since some of the positivity constraints will be violated. This situation would correspond to flowing into the interior of the allowed region, so that we are no longer at the boundary. We illustrate this case in figure \ref{fig:sings}.
The solution is that if we want to remain at the boundary, we must adopt the zero as a new vector in the solution to crossing, with zero OPE coefficient. Only then we proceed with the flow, which leads to a positive OPE coefficient for that operator. In this way we have gone from a solution with $K$ vectors to one with $K+1$. 

Since the counting of degrees of freedom has changed, we must deal with the construction of the functional. In particular, the assignment of vectors as doubled or singles depends on $K$. In practice, the resolution is simple. At the singularity, we take some doubled vector and turn it into a single, and then add the new vector $\vv *$, also as a single. For instance,
\bea
&&\bphi(\bullet)\propto \langle \mbf f_1\,\ldots \mbf f_{n_f}\,\mbf s_1\,\ldots \mbf s_{n_s}\, \mbf d_1\,\partial_\Delta \mbf d_1\ldots \mbf d_{n_d}\,\partial_\Delta \mbf d_{n_d}\, \bullet\rangle\nonumber \\
\to&& \bphi'(\bullet) \propto \langle \mbf f_1\,\ldots \mbf f_{n_f}\,\mbf s_1\,\ldots \mbf s_{n_s} \mbf d_{n_d}\, \mbf v^*\, \mbf d_1\,\partial_\Delta \mbf d_1\ldots \mbf d_{n_{d-1}}\,\partial_\Delta \mbf d_{n_{d-1}}\, \bullet\rangle \nonumber
\eea
This is a consistent thing to do, since at the singularity we have $\bphi(\mbf v^*)=\bphi(\partial_\Delta \mbf v^*)=0$. In particular, the linear dependencies guarantee that after the swap we will still have $\bphi'(\partial_\Delta \mbf d_{n_d})=\bphi'(\partial_\Delta \mbf v^*)=0$, so that these vectors are indeed singles satisfying the tangency conditions. In fact, at the singularity the two functionals $\bphi, \bphi'$ are actually identical. In particular, notice that the functional will be continuous across this transition, but not its first derivative. Accordingly, this singularity leads to a kink not in a bound itself, but in its first derivative, as follows from equations~\reef{eq:opemax1} and~\reef{eq:gapmax1}.

\subsubsection*{2. One operator less}

This singularity is the counterpart of the one above when one is flowing from the opposite direction. In this case, one would see an operator's OPE coefficient going to zero. If we would continue flowing, the OPE coefficient would become negative and we would violate unitarity and positivity. This would correspond to a flow that goes above the boundary. In order to preserve unitarity, we must remove the operator from the spectrum, so that the solution goes from $K$ to $K\!-\!1$ vectors. Since we have a lost a vector, we must now do something about the functional. We can always choose the decoupling vector to be a single, since the choice of singles and doubles is arbitrary. So the solution is simply to delete that single from the solution (and the functional), and turn some other single into a double, the exact converse of what we did above.

\subsection*{3. Kinks}

Finally, let us try to understand kinks. A kink is a discontinuity in the first derivative of a bound, and from equations \reef{eq:opemax1},\reef{eq:gapmax1} this can happen if the functional itself is discontinuous. To see how this can occur, notice that in case 2 above we made an important assumption, which is that at the point of the decoupling of the operator, there are at least two singles in the solution to crossing. If this is not the case, then our resolution fails. At the singularity, the solution to crossing becomes the most extremal possible, containing only doubles and fixed vectors. In other words, the solution is in some sense the sparsest it can be. We claim that this is exactly what happens at kinks. 

More pragmatically, we need to understand how the flow can proceed. The solution must be that beyond the singularity a new single appears. In other words, a kink corresponds to a swapping of two singles. Geometrically, one is moving across a higher codimension boundary of parameter space, across which the functional can jump discontinuously. In practice, to continue with the flow we must find this new single. How to do it? Well, we simply replace the decoupling single with some other vector $\mbf v^*$ and demand that the tangency condition is satisfied:
\bea
\bphi'_{\mbf v^*}(\partial_\Delta \mbf v^*)=0
\eea
This should be thought of as an equation for $\mbf v^*$. It is possible that there could be several solutions to this equation, in which case we must pick the one for which the functional is positive everywhere. This equation can be solved efficiently, and the details will be presented elsewhere \cite{futurework}.

\subsubsection{A $D=1$ example}
\label{sec:singres}
\begin{figure}
\begin{center}
\begin{tabular}{cc}
\includegraphics[width=7.45cm]{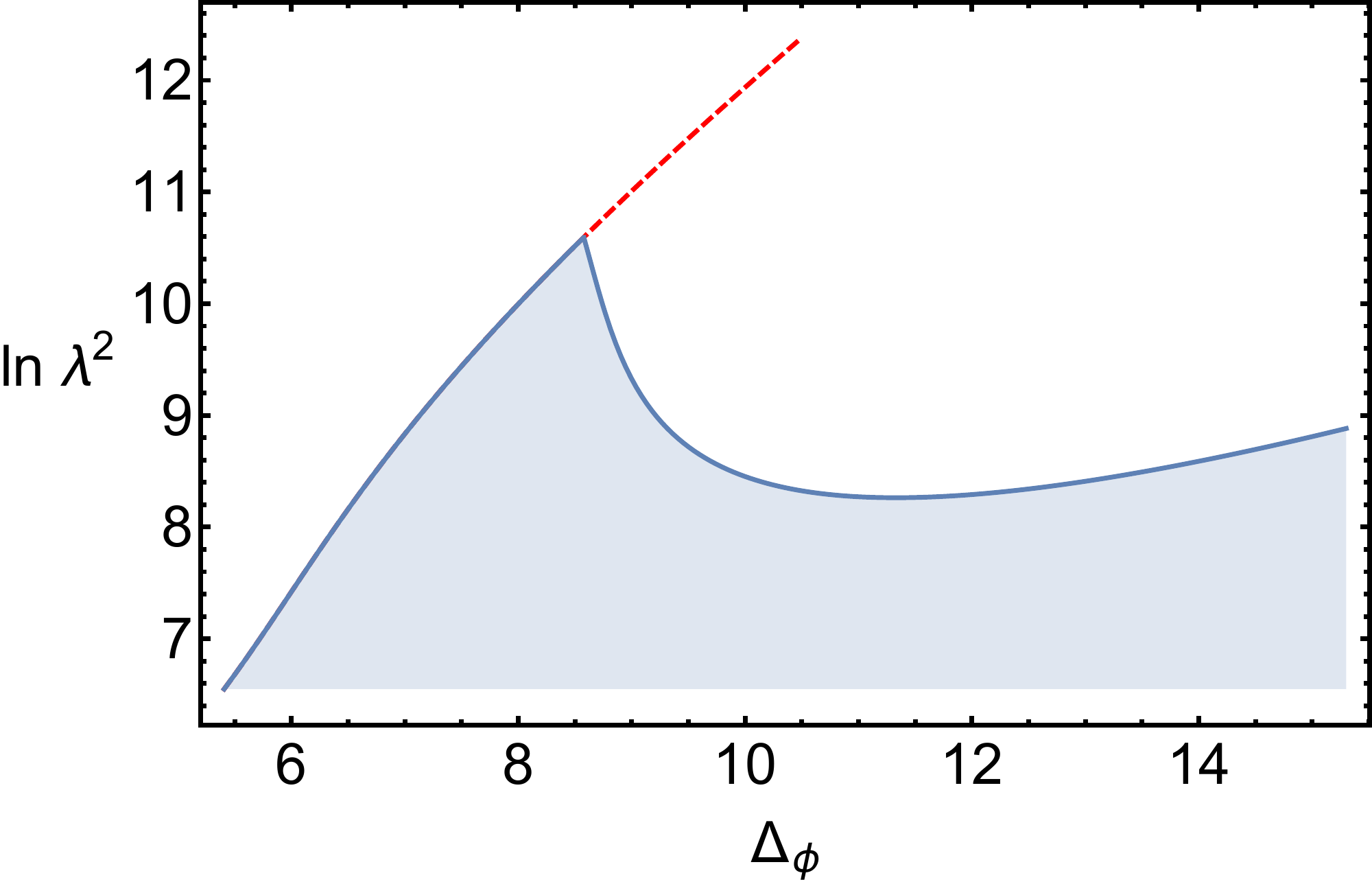}
&
\includegraphics[width=7cm]{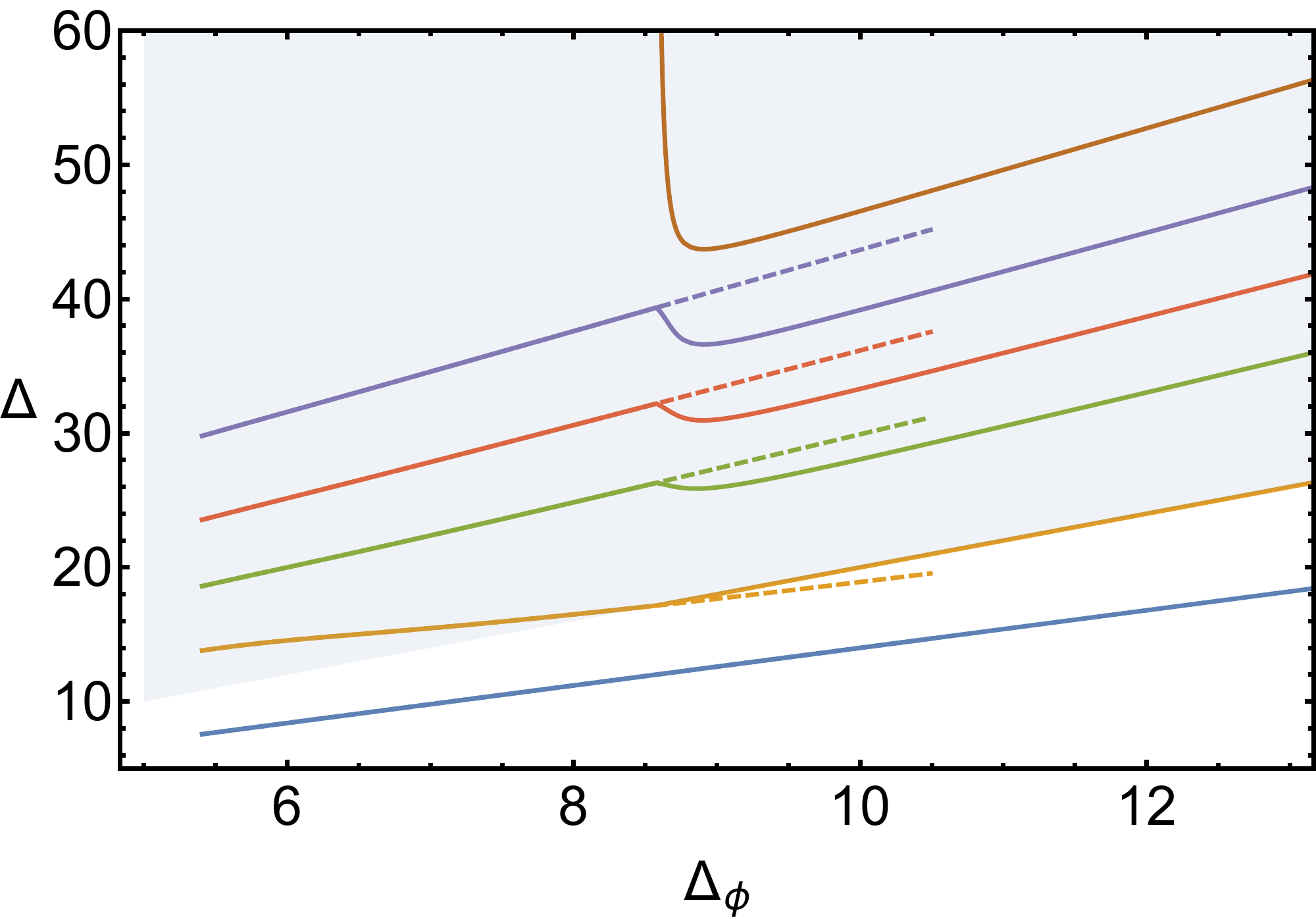}
\end{tabular}
\end{center}
\caption{A human-made kink. On the left, bound on the OPE coefficient of an operator of dimension $\Delta=1.4\,\Delta_\phi$, with all other operators above $\dgap=2\,\Delta_\phi$. On the right the corresponding spectra. Only to the right of the kink is there an operator saturating the gap. To its left, an extra operator at $\Delta=\infty$ makes a finite contribution to the solution. The dashed lines are the continuation of the extremal solution ignoring $\dgap$. All computations done with $N=10$ components.}
\label{fig:sing}
\end{figure}

As an example, let us show how a singularity can arise even in $D=1$. For definiteness, we consider maximizing the OPE coefficient of an operator with dimension $\Delta_*=1.4 \Delta_\phi$, and with a gap $\dgap=2\Delta_\phi$. The resulting bound on the OPE coefficient is shown in figure \ref{fig:sing}.
To understand the origin of the shark's fin, it is easiest to look at the spectrum of the solution. When flowing from the left, the spectrum does not contain an operator with $\Delta=\dgap$. In this way the gap could be varied without affecting the bound. However, at some point the first operator above the gap collides with it. In terms of the flow, this operator must now become fixed. Fixing it loses a degree of freedom which we must get from somewhere else. The answer is that a fixed operator at $\Delta=\infty$ simultaneously becomes unfixed, and comes down very rapidly. Flowing from the right the situation is reversed. Finally, were the gap not present we could have simply continued the flow. This would be the analog of flowing into a non-unitary region in a more complicated setup.

\section{Conclusions and Outlook}
In this paper, we have introduced a new bootstrap technique for determining numerical approximations to the spectra of conformal field theories. This technique is based on the concept of extremality: CFT correlation functions that saturate bounds have sparser spectra and satisfy a set of extremality equations. Once a reasonable first approximation to a solution of the crossing equations is found, these equations can be used to improve this approximation very efficiently. In fact, they can even be used to construct such solutions from scratch, as in our upgrading example of section \ref{sec:upgrade}. They can also be used to solve for the variation of the spectrum under smooth deformations. We have shown that this works extremely well, leading to drastic improvements in computational efficiency of up to several orders of magnitude. A public version of the extremal flows code will be available in the future as part of the {\tt JuliBootS} package \cite{Paulos2014a}. Preliminary versions are available upon request.

In this note we have focused on one-dimensional applications. Our motivation for this was simplicity: in a single correlation function in a 1d CFT the operators appearing in the OPE are labelled by only one continuous parameter. This allows us to focus purely on continuous flows. In higher spacetime dimensions for instance, operators will be labeled also by the discrete spin quantum number. One expects then that along the flow there could be discrete transitions, where operators swap spin. The flow is non-analytic in this case, and requires more care, as outlined in section \ref{sec:sings}. We have explicitly checked that there is no obstacle to generalizing our methods to these cases, and that these non-analyticities can be dealt with in a straightforward (and efficient) manner. We hope to report on these results in the near future \cite{futurework}.

An important application of our methods is to the bootstrap of multiple correlation functions \cite{Kos2016,Kos2015,Kos:2014bka}. In this case, the state-of-the-art technique currently relies on semidefinite programming \cite{Poland:2011ey,Simmons-Duffin2015}. In this approach, one does not construct a solution to crossing directly. Rather, in a disallowed region of parameter space one has a positive functional, essentially the same as the one used throughout this paper. From the functional it should be possible to extract the spectrum, and once we have this we can use our methods straightforwardly. This should be extremely useful, since the multiple correlator bootstrap is computationally expensive. Our results suggests that one should concentrate one's energies on obtaining a single initial point on a bound as efficiently as possible, since flowing to nearby solutions is then significantly cheaper.  Obviously, it would be even better if we could apply our ``upgrading'' approach to multiple correlators, thereby sidestepping this rather expensive first step completely.

The results of this paper show, as advocated in \cite{ElShowk:2012hu,El-Showk2014a}, that the approximate solution to crossing symmetry obtained for extremal CFTs contains valuable information that is almost invisible from the point of view of bounds. For instance, there can be singularities which do not lead to kinks in bounds. Such singularities could very well signal the presence of interesting CFTs, which would be completely missed in a simplistic examination of bounds. While spectrum extraction remains the best possibility, a poor man's alternative would be at the very least to look not only at bounds but their derivatives in the search for kinks. Unfortunately good results require a fairly decent resolution, and hence more computational power, which suggests ours is a better approach.

In section~\ref{sec:det} we have related the method of determinants of Gliozzi~\cite{Gliozzi2013} to the usual bootstrap approach, as well as to the results of this paper. We have argued that if this method is to become systematic, one needs to add extra constraints. A possibility is that these constraints are nothing but the extremality conditions that we have proposed for unitary CFTs. As suggested in the same section, the most promising possibility is to pursue unitary flows into non-unitary regions. As a concrete example, in the two dimensional bootstrap, it has been observed that a bound on dimension of the leading scalar operator appearing in an OPE of two other identical scalars has a sharp kink at the location of the 2d Ising model~\cite{Rychkov:2009ij}. This special point lies on a line of exact solutions to crossing symmetry which can be written down analytically. The origin of this kink was understood in~\cite{El-Showk2014a} as arising precisely from the decoupling of certain operators at the Ising point, meaning that some OPE coefficients become zero. Now, we simply remark that the line of exact solutions makes sense even below the Ising point, whereupon they become non-unitary. In fact, such a line eventually terminates at the Lee-Yang singularity. It is natural to conjecture then that flowing past the Ising point will take us along this line of non-unitary solutions.

The Ising point itself is of considerable interest, especially in three dimensions. In particular, our methods promise to greatly improve the accuracy of previous work \cite{El-Showk2014a}. One natural conjecture is that one can define the Ising kink as the point where the spectrum is sparsest, containing only fixed and doubled vectors, in the nomenclature of section \ref{sec:extflow}. This should allow us to pinpoint very accurately the Ising point for any truncation. We hope to test this conjecture in the near future.

\acknowledgments{
We are grateful to J. Penedones, V. Rychkov and A. Vichi for discussions. MFP is supported by a Marie Curie Intra-European Fellowship of
the European Community's 7th Framework Programme under contract number PIEF-GA-2013-623606.}

\appendix
%

\section{Error bound on derivatives}\label{sec:errder}
Here we will estimate the convergence of the crossing symmetry sum rule,
\bea
\vv 0+\sum_i a_i \vv i=0
\eea
Recalling the definition of the $a_i$ and the vectors $\vv i$, a particular component of this equation takes the form
\bea
\sum_{\Delta,l} (\lambda_{\Delta,l})^2
\left(\frac{\partial}{\partial u}\right)^m 
\left(\frac{\partial}{\partial v}\right)^n
F_{\Delta,l}^{(\phi)}(u,v)\bigg |_{u=v=1/4}=0
\eea
Let us do a change of coordinates to a more convenient set. We will use radial quantization coordinates, setting $x_1^\mu =0,x_4^\mu=\infty$ and $x_2^\mu=r_2 n_2^\mu, x_3^\mu=r_3 n_3^\mu$ with the coordinates on the sphere $n_2^2=n_3^2=1$. In terms of these we have
\bea
u=\left(\frac{r_2}{r_3}\right)^2=e^{-2\beta}, \qquad v=(1-e^{-\beta})^2+2(1-\alpha), \qquad \alpha\equiv n_2\cdot n_3
\eea
The crossing symmetric point becomes $\beta=\ln 2$ and $\alpha=1$. From the definition
\bea
F_{\Delta,l}^{(\phi)}(u,v)=v^{\Delta_\phi} G_{\Delta,l}(u,v)-u^{\Delta_\phi} G_{\Delta,l}(v,u),
\eea
follows that we must take odd derivatives with respect to $\beta$. Hence we can consider instead
\bea
v^{\Delta_\phi}\sum (\lambda_{\Delta,l})^2 
\left(-\frac{\partial}{\partial \beta}\right)^{2m-1}
\left(\frac{\partial}{\partial \alpha}\right)^n
G_{\Delta,l}(u,v)\bigg |_{\beta=\ln 2, \alpha=1}+\mbox{lower derivatives}=0
\eea
We want to estimate the convergence of the terms with highest derivatives, which is now seen to be equivalent to studying the convergence of derivatives of the four-point function of $\phi$. Hence, let us focus on the latter.

The convergence of the four-point function itself at general $u,v$ has been studied before \cite{Pappadopulo2012}. As we will see, their argument generalizes straightforwardly when we consider derivatives of the four-point function instead. We begin by writing the four point function in radial quantization as
\bea
\mathcal L(\beta,\alpha)&=&r_2^{\df} r_3^{\df} \langle \phi|\phi(x_3)\phi(x_2)|\phi\rangle=\sum_{\Delta,l} (\lambda_{\Delta,l})^2 \sum_{k=0}^{+\infty} \langle \cO_{\Delta,l},k, n_3|\cO_{\Delta,l},k, n_2\rangle\, e^{-\beta E_{\cO,k}} 
\eea
In the expression above, the matrix elements correspond to level $n$ descendant states of primaries with dimension $\Delta_{\mathcal O}$. The energies are $E_{\cO,k}=\Delta_\cO+k$. The matrix elements depend on $n_2,n_3$ only through their internal product; in fact they are related to Gegenbauer polynomials:
\bea
\langle \cO,k,n_2|\cO,k,n_3\rangle =\sum_{p\leq k} c_p\, C^{\left(\frac{d-2}2\right)}_p(\alpha), \qquad c_p\geq 0.
\eea
We are interested in placing a bound on contributions to the expression above with $\Delta\geq \dels$, for $\dels$ sufficiently large. Such contributions are bounded from above by the case $\alpha=1$. We can then write:
\bea
\mathcal L(\beta,\alpha=1)\equiv \mathcal L(\beta)
&=&\int_0^{+\infty} \ud E\, f(E) e^{-\beta E}, \qquad f(E)\geq 0 \nonumber \\
\mathcal L(\beta,\dels)&=&\int_{\dels}^{+\infty} \ud E f(E) e^{-\beta E}.
\eea
To get a bound on $\mathcal L(\beta,\dels)$ the basic argument follows from the Hardy-Littlewood tauberian theorem. Using the OPE when $x_2\simeq x_3$ implies
\bea
\lim_{\beta\to 0} \mathcal L(\beta,1)=\frac{1}{\beta^{2\Delta_\phi}}.
\eea
The theorem then implies that in the same limit $f(E)\simeq E^{2\Delta_\phi}$. After some work this allows one to find:
\bea
\mathcal L(\beta,\dels)\leq \mathcal N\, (\dels)^{2\Delta_\phi}\, C^{\dels}, \qquad C<1, \dels\to \infty.
\eea
for some normalization constant $\mathcal N$ independent of $\dels$. This proves that contributions above sufficiently large $\dels$ fall off exponentially fast. We now want to make a similar statement for derivatives of $\mathcal L(\beta,\dels)$. Consider first acting with derivatives with respect to $\alpha$. We have
\bea
\left(\frac{\partial}{\partial \alpha}\right)^n C^{\left(\frac{d-2}2\right)}_k(\alpha)\bigg|_{\alpha=1}=2^k\frac{\left(\frac{d-2}2\right)_n\,\Gamma\left(d-2+k+n\right)}{\Gamma\left(d-2+2n\right)\,(k-n)!}>0
\eea
We see that these derivatives act on the radial quantization elements in such a way that positivity is preserved. Hence we can write
\bea
\left(\frac{\partial}{\partial \alpha}\right)^n \mathcal L(\beta,\alpha)\bigg|_{\alpha=1}=\int_0^{+\infty} \ud E\, f^\alpha_n(E) e^{-\beta E}
\eea
for some $f_n^\alpha(E)\geq 0$. We can consider derivatives with respect to $\beta$ in the same way:
\bea
\left(-\frac{\partial}{\partial \beta}\right)^m \mathcal L(\beta,\alpha)\bigg|_{\alpha=1}=\int_0^{+\infty} \ud E f_m^{\beta}(E) e^{-\beta E}\equiv
\eea
with $f_m^{\beta}(E)\equiv E^m f(E)\geq 0$. Combinations of derivatives clearly lead to similar representations involving some positive function of the energy.
Now, as $\beta \to 0$ we have $x_2\to x_3$ and hence:
\bea
&&\lim_{\beta\to 0} \mathcal L(\beta,\alpha)=\frac{1}{\left[\beta^2+2(1-\alpha)\right]^{\df}}\nonumber\\
&\Rightarrow& \lim_{\beta\to 0}
\left(-\frac{\partial}{\partial \beta}\right)^m
 \left(\frac{\partial}{\partial \alpha}\right)^n \mathcal L(\beta,\alpha)\bigg|_{\alpha=1}
=\frac{2^n(\df)_{n}(2\df+2n)_m}{\beta^{2\df+m+2n}}
\eea
At this point the argument of \cite{Pappadopulo2012} goes through unmodified. In particular, the contribution of operators of dimension $\Delta\geq \Delta_*$ is bounded as
\bea
\left(-\frac{\partial}{\partial \beta}\right)^{2m-1}
 \left(\frac{\partial}{\partial \alpha}\right)^n \mathcal L(\beta,\dels)\leq \mathcal N_{m,n} (\Ds)^{2\df+2m+2n-1} C^{\dels},
\eea
which is the desired result.
\bibliography{Biblio}
\bibliographystyle{JHEP}

\end{document}